\documentclass[twocolumn,showpacs,preprintnumbers,amsmath,amssymb,superscriptaddress]{revtex4}
\usepackage{amssymb}
\usepackage{graphicx,color}
\usepackage{bm}

\begin{document}
\title{Effects of initial flow velocity fluctuation in event-by-event (3+1)D hydrodynamics}

\author{Longgang Pang}
\affiliation{Nuclear Science Division, MS 70R0319, Lawrence Berkeley National Laboratory, Berkeley, CA 94720}
\affiliation{Interdisciplinary Center for Theoretical Study and Department of Modern Physics, University of Science and Technology of China, Hefei 230026, China}

\author{Qun Wang}
\affiliation{Interdisciplinary Center for Theoretical Study and Department of Modern Physics, University of Science and Technology of China, Hefei 230026, China}

\author{Xin-Nian Wang}
\affiliation{Nuclear Science Division, MS 70R0319, Lawrence Berkeley National Laboratory, Berkeley, CA 94720}
\affiliation{Institute of Particle Physics and Key Laboratory of Quarks and Lepton Physics, Central China Normal University, Wuhan 430079, China}

\begin{abstract}
Hadron spectra and elliptic flow in high-energy heavy-ion collisions are studied within a (3+1)D ideal hydrodynamic model with fluctuating initial conditions given by the AMPT Monte Carlo model. Results from event-by-event simulations are compared with experimental data at both RHIC and LHC energies.
Fluctuations in the initial energy density come from not only the number of coherent soft interactions of 
overlapping nucleons but also incoherent semi-hard parton scatterings in each binary nucleon collision. Mini-jets from semi-hard parton scatterings are assumed to be locally thermalized through a Gaussian smearing and give rise to non-vanishing initial local flow velocities. Fluctuations in the initial flow velocities lead to harder transverse momentum spectra of final hadrons due to non-vanishing initial radial flow velocities. Initial fluctuations in rapidity distributions lead to 
expanding hot spots in the longitudinal direction and are shown to cause a sizable reduction of final hadron elliptic flow at large transverse momenta. 

\end{abstract}

\pacs{12.38.Mh,24.10.Nz,25.75.-q,25.75.Ld}

\maketitle

\section{Introduction}

One of the striking phenomena in high-energy heavy-ion collisions at the Relativistic Heavy-ion Collider (RHIC) and the Large Hadron Collider (LHC)  is the collective flow \cite{Ackermann:2000tr,Adams:2004bi,Adler:2003kt,Afanasiev:2009wq,Back:2004mh,Aamodt:2010pa,collaboration:2011hfa} generated by the tremendous pressure of the
dense matter, from the quark-gluon plasma in the early time to the hadronic resonance gas in the late stage of the evolution.
Such collective transverse flow influences not only the transverse momentum spectra but also the azimuthal anisotropy or 
anisotropic flow of the final hadrons \cite{Ollitrault:1992bk, Voloshin:1994mz,Kolb:2000sd}.
Anisotropic flow arises from the collective expansion of dense matter with initial geometric anisotropy. The simplest example
is the hydrodynamic expansion of dense matter with a smooth initial energy density distribution in non-central heavy-ion collisions. 
The initial radial pressure gradient in the almond-shaped dense matter is asymmetric in the azimuthal angle.
Such an asymmetric pressure gradient drives the system into a transverse expansion and 
transforms the initial geometric asymmetry into momentum asymmetry in azimuthal angle.
The second Fourier coefficients of hadron azimuthal distributions are known as the elliptic flows. 
The large values of elliptic flow as measured in semi-central heavy-ion collisions 
at RHIC and LHC suggest the formation of a strongly coupled quark-gluon plasma close to a perfect 
fluid \cite{Huovinen:2001cy,Heinz:2001xi,Gyulassy:2004zy,Jacobs:2004qv,Muller:2006ee, Shuryak:2008eq, Heinz:2008tv}.
Comparisons of experimental data on elliptic flow and viscous hydrodynamic simulations can now provide 
phenomenological constraints on the specific shear viscosity (the ratio of shear viscosity to entropy density)
of the quark-gluon plasma \cite{Teaney:2003kp,Romatschke:2007mq,Song:2007ux,Dusling:2007gi,Song:2010mg}. 

One of the critical inputs for the hydrodynamic model of heavy-ion collisions is the initial condition. Smoothed distributions
of the initial energy density from either a Glauber or Color Glass Condensate (CGC) model of parton production were
used in some recent hydrodynamic calculations \cite{Kolb:2000sd, Romatschke:2007mq,Song:2007ux,Dusling:2007gi,Song:2010mg,Hirano:2005xf}. 
However, one has to resort to event-by-event hydrodynamic simulations \cite{Steinheimer:2007iy,Werner:2010aa,Andrade:2010xy,Petersen:2010cw,Holopainen:2010gz,Schenke:2010rr,Albacete:2011fw, Schenke:2011bn, Qiu:2011iv, Qiu:2011hf, Schenke:2012wb, Gardim:2012yp}
to take into account the fluctuation in the initial conditions. Such fluctuating initial conditions have been shown to be responsible
for odd harmonic flows (harmonic coefficients of the azimuthal anisotropy), such as the 
triangular flow \cite{Alver:2010gr}, as well as the double-peak structure of dihadron 
correlation \cite{Adler:2005ee,Adare:2008ae,Abelev:2008ac,Agakishiev:2010ur,Alver:2009id,ALICE:2011ab,Aad:2012bu} in 
the final hadron spectra. These odd harmonic flows and dihadron
correlations persist even in the most central heavy-ion collisions due to fluctuation of the initial local parton density \cite{Ma:2010dv}.
Because of approximate longitudinal boost invariance of the local parton density in the initial condition, anisotropic flows
in transverse momentum spectra are also correlated in pseudo-rapidity leading to the observed ``ridge'' 
structure in dihadron correlation in azimuthal angle and 
pseudo-rapidity \cite{Aad:2012bu,Ma:2010dv,Adams:2004pa,Putschke:2007mi,Abelev:2009af,Takahashi:2009na, Werner:2011xr, Xu:2011fe}.

Most of recent event-by-event hydrodynamic studies \cite{Holopainen:2010gz,Schenke:2010rr,Albacete:2011fw, Schenke:2011bn,Qiu:2011hf, Schenke:2012wb,Gardim:2012yp,Chaudhuri:2011pa} employ the Glauber  \cite{Miller:2007ri} or CGC 
model  \cite{Drescher:2006pi, Drescher:2006ca} of parton production for the initial transverse energy density distribution. 
The Monte Carlo (MC) Glauber model assumes an initial energy density that is proportional to the transverse density of the number of 
wounded nucleons or a linear combination of the number of wounded nucleons and binary nucleon-nucleon collisions. 
The Monte Carlo (MC) CGC inspired models use the Kharzeev-Levin-Nardi (KLN) description \cite{Kharzeev:2002ei, Kharzeev:2004if} of 
initial gluon production per wounded nucleon pair whose produced gluon multiplicity also depends on the impact parameter.
Such fluctuating or bumpy initial energy density distributions in the event-by-event hydrodynamic simulations affect both the
transverse momentum spectra and the azimuthal anisotropic flow as compared to the event-averaged smooth initial conditions,
due to the increased local pressure gradient around hot spots or cold valleys. The transverse expansion of these hot spots
amid the overall expanding medium also leads to large values of odd harmonic azimuthal anisotropy of the final hadron
spectra which will give rise to a conic structure in the dihadron azimuthal correlation \cite{Ma:2010dv,Takahashi:2009na,Werner:2011xr}.

The initial conditions used for recent event-by-event hydrodynamic simulations
have mostly assumed zero local flow velocities. Some of the hydro 
simulations \cite{Steinheimer:2007iy,Werner:2010aa,Andrade:2010xy} have included 
local velocity and longitudinal fluctuation. But their effects have not been systematically studied before.
Since initially produced partons are basically uncorrelated 
in different nucleon-nucleon collisions, such an assumption is approximately correct for initial conditions 
that are smoothed over a transverse area much larger than the nucleon size.
However, when fluctuation of initial energy density over the range of a nucleon size or smaller is considered, the initial 
local flow velocities are nonzero and their effects are non-negligible. The nonzero local flow 
velocity can arise from multiple parton correlation in mini-jets production which 
becomes the dominant mechanism for initial parton production in high-energy heavy-ion collisions at RHIC 
and LHC \cite{Wang:1991hta, Gyulassy:1994ew, Wang:2009qb}. Mini-jets are clusters of many partons collimated in phase space. 
After initial thermalization, correlations of these partons associated with a pair of mini-jets are not necessarily
destroyed and thus lead to non-vanishing local flow velocities. Such local flow velocities
are expected to increase the final hadron multiplicity and the slope of hadron transverse momentum spectra.
They should also lead to small initial collective radial flow velocity at the outer region of the dense matter, which
was found empirically important to explain the experimental data on HBT correlations \cite{Pratt:2005bt}. The initial local
transverse flow velocities due to mini-jets can also lead to intrinsic same-side and away-side dihadron correlations
which are not induced by the collective expansion of the dense matter. Similarly, fluctuations in the local longitudinal
flow velocity are also important and should be included in (3+1)D hydrodynamics \cite{Florchinger:2011qf}.

In this paper, we will study the effects of fluctuating initial flow velocities  within an ideal (3+1)D 
hydrodynamic model using the initial conditions as provided by the AMPT (A Multi-phase Transport ) 
Monte Carlo model \cite{Zhang:1999bd}. Such initial conditions contain mini-jets
given by the HIJING Monte Carlo model \cite{Wang:1991hta, Gyulassy:1994ew} and initial
thermalization via the parton cascade within the AMPT model. 
We will study the effects of initial local flow velocity fluctuations in both transverse and longitudinal directions
on the final hadron multiplicity distribution, transverse momentum spectra, elliptic flow 
as well as dihadron correlations. We will show the importance of the initial local flow velocity fluctuation in
the study of hadron spectra and elliptic flow and therefore the current effect of extracting shear viscosity
from comparisons between hydro calculations and experimental data.

The rest of this paper is organized as follows. In Sec. II, we will give a brief description of the (3+1)D ideal 
hydrodynamic model that we have developed, including a new projection method for calculation of the freeze-out
hyper surface and determination of initial conditions from the AMPT model. We will compare
the calculated hadron spectra and elliptic flow with experimental data from RHIC and LHC in
Sec. III. We will investigate in detail the effects of initial local flow velocities on hadron spectra and elliptic
flow by comparing hydrodynamic simulations with and without initial local flow velocities in Sec. IV. 
We study the sensitivity of final hadron spectra and elliptic flow in Sec. V with a comparison between hydro results
with a Chemical Equilibrated (CE) Equation Of State(EoS) and Partial Chemical Equilibrium (PCE) EoS.
We conclude in Sec. VI with a discussion on the dihadron correlations from event-by-event hydrodynamics. 
We will give a brief description of the SHASTA algorithm that we use to solve the (3+1)D hydrodynamic
equations in the Appendix.

\section{Ideal (3+1)D hydrodynamics}

\subsection{Conservation equations}

The ideal hydrodynamic model of high-energy heavy-ion collisions is based on the assumption that local thermal
equilibrium is achieved at some initial time $\tau_{0}$ and the evolution of the system afterwards can 
be described by conservation equations for energy-momentum tensor and net baryon current,
\begin{eqnarray}
\partial_{\mu}T^{\mu\nu}&=&0,\\
\partial_{\mu} J^{\mu}&=&0,
\end{eqnarray}
where the energy-momentum tensor and net baryon current can be expressed as
\begin{eqnarray}
T^{\mu\nu}&=&(\varepsilon+P)u^{\mu}u^{\nu}-Pg^{\mu\nu}, \nonumber \\ 
J^{\mu}&=&nu^{\mu},
\label{eq:hydro}
\end{eqnarray}
in terms of the local energy density $\varepsilon$, pressure $P$,  the metric tensor $g^{\mu\nu}$, net baryon density $n$ 
(or any conserved charges) and  time-like 4-velocity $u^{\mu}$ with $u^2=1$.
The above 5 equations contain 6 variables and can be closed by the equation of state (EoS) $P=P(\varepsilon,n)$.
We will use the parameterized EoS s95p-v1 by Huovinen and Petreczky \cite{Huovinen:2009yb} which has a cross-over 
between the lattice QCD results at high temperature and hadron resonance gas  below the cross-over temperature. The
chemical freeze-out temperature in this parameterization is set at 150 MeV. 
Such an EoS is valid only for zero baryon number density or chemical potential. For simplicity,
however, we will assume the same EoS for all region of dense matter in our calculation, including large 
rapidity region where the net baryon density is nonzero.

The velocity of a fluid element in Cartesian coordinates $x^{\mu}=(t,x,y,z)$ is defined as \cite{weinberg}
\begin{equation}
  u^{\mu} \equiv \frac{d x^{\mu}}{d\sigma}
  \equiv u_{0}(1,\vec{\tilde{v}}_{\perp},\tilde{v}_{z})
  \label{eq:flow_tz}
\end{equation}
where $\sigma = \sqrt{t^2 - x^2 - y^2 -z^2}$ and spatial components of the flow velocity are 
defined as $\tilde{v}_{i} = u^{i}/u_{0}$ ($i=x,y,z$). The time-component is $u_{0}=1/\sqrt{1-\tilde{v}^2}$.

We will work in this paper with the invariant-time coordinate $X^{\mu}=(\tau,x,y,\eta_{s})$ where $\tau = \sqrt{t^{2}-z^{2}}$
and the spatial rapidity $\eta_{s}$ are defined as,

\begin{eqnarray}
  t &=& \tau \cosh\eta_{s}, \nonumber \\
  z &=& \tau \sinh\eta_{s} \; .
\end{eqnarray}
The metric tensor $g^{\mu\nu}=\mathrm{diag}(1,-1,-1,-1/\tau^{2})$ and correspondingly $g_{\mu\nu}=\mathrm{diag}(1,-1,-1,-\tau^{2})$
are given by the invariant line element $ds^{2} = g_{\mu\nu}dX^{\mu}dX^{\nu} = d\tau^2 -dx^2 -dy^2 - \tau^2 d\eta_{s}^{2}$ in
the invariant-time coordinate. The velocity 4-vector in this coordinate is,
 \begin{eqnarray}
  U^{\mu} &\equiv & \frac{dX^{\mu}}{d\sigma} = \frac{dX^{\mu}}{dx^{\nu}}\frac{dx^{\nu}}{d\sigma} 
  = \frac{dX^{\mu}}{dx^{\nu}} u^{\nu}  \\
  &=& \left( 
  \begin{array}{l}
	u^{0}\cosh\eta_s - u^{z}\sinh\eta_s \\
	\vec u_{\perp} \\
	\frac{1}{\tau}(-u^{0}\sinh\eta_s + u^{z}\cosh\eta_s )
  \end{array} \right)
  \equiv U^{\tau}\left(
  \begin{array}{l}
	1\\
	\vec v_{\perp} \\
	\frac{v_{\eta}}{\tau}
	\nonumber
  \end{array} \right)
  \label{eq:Utau}
\end{eqnarray}
where $v_{z}$, $\vec v_{\perp}$ and $v_{\eta}$ are defined as \cite{Hirano:2001eu},
 \begin{eqnarray}
  \vec v_{\perp} &=&\vec {\tilde{v}}_{\perp}\cosh(y_{v})/\cosh(y_{v}-\eta_{s}),\nonumber \\
  v_{\eta} &=& \tanh(y_{v}-\eta_{s}), \nonumber 
\end{eqnarray}
$y_{v}$ denotes the rapidity of the longitudinal flow velocity as given by $\tilde{v}_{z} = \tanh y_{v}$ and 
 $U^{\tau} = 1/\sqrt{1-v_{\perp}^{2}-v_{\eta}^{2}}$.

Since we assume an EoS that is independent of the local baryon number density or chemical potential, 
the baryon current conservation equation decouples from the energy-momentum conservation. We will
regard $T^{\tau\tau}$, $T^{\tau x}$, $T^{\tau y}$, $T^{\tau \eta}$ and $J^{\tau}$ as independent variables
in the conservation equations. Other components of the energy-momentum tensor and the baryon current
can be expressed in terms of these $5$ variables from the definitions in Eq.~(\ref{eq:hydro}).
For example, from definition $T^{\tau x} = (\varepsilon + P)U^{\tau}U^{x}$, one can express
$T^{x x} = (\varepsilon+P)U^{x} U^{x} + P = v_{x} T^{\tau x} + P$.  With these independent variables,
we can use the SHASTA(SHarp And Smooth Transport Algorithm) algorithm, which is designed to 
solve partial differential equations with the form $\partial_{t}(T) + \partial_{i}(v_{i}T) = S$,  to solve 
the hydrodynamic equations. These conservation equations can be cast in the following 
form by variable substitutions,
\begin{eqnarray}
  \partial_{\tau}(\tau T^{\tau\tau})-\tau \mathbf{\nabla}\cdot(\mathbf{v} T^{\tau\tau}) & = & S^{\tau}, \nonumber\\
    \partial_{\tau}(\tau \vec T^{\tau \perp})- \tau \mathbf{\nabla}\cdot(\mathbf{v} \vec T^{\tau \perp})& = & \vec S^{\perp}, \nonumber\\
    \partial_{\tau}(\tau T^{\tau\eta})- \tau \mathbf{\nabla}\cdot(\mathbf{v} T^{\tau \eta})& = & S^{\eta}, \nonumber\\
    \partial_{\tau}(\tau J ^{\tau})- \tau \mathbf{\nabla}\cdot(\mathbf{v} J^{\tau }) & = & 0, 
    \label{eq:hydro2}
\end{eqnarray}
with the source terms,
\begin{equation}
  \left( \begin{array}{c}
	S^{\tau}\\
	\vec S^{\perp}\\
	S^{\eta}
	\end{array} \right)
	=
	\left( \begin{array}{l}
	  \tau \mathbf{\nabla}\cdot (\mathbf{v} P)-v_{\eta}^2(T^{\tau\tau}+P)-P \\
       -\tau \vec\partial_{\perp} P \\
	   -(1/\tau)\partial_{\eta} P -2 T^{\tau\eta}
	\end{array} \right),
   \label{eq:source_term}
\end{equation}
where $\mathbf{\nabla}\cdot(\mathbf{v} R)=\vec \partial_{\perp}\cdot (\vec v_{\perp} R)+\partial_{\eta_{s}}(v_{\eta}R)/\tau$
for any variable $R$. The energy density is determined from $T^{\tau \nu}$ through a root finding method by iterating the 
following equation,
\begin{equation}
  \varepsilon = T^{\tau\tau} - \frac{M^2}{T^{\tau\tau}+P(\varepsilon)},
  \label{eq:root_finding}
\end{equation}
 to an accuracy $\left|\delta \varepsilon\right|<10^{-15}$,
 where $M^{2}=(T^{\tau \perp})^2+(\tau T^{\tau \eta})^2$.
 The initial value of $\varepsilon$ for the iteration is approximated by $\varepsilon=T^{\tau\tau}$.
 The flow velocity is given by
\begin{eqnarray}
\vec v_{\perp}&=&\vec T^{\tau \perp}/[T^{\tau\tau}+P(\varepsilon)], \;\; \\
v_{\eta}&=&\tau T^{\tau \eta}/[T^{\tau\tau}+P(\varepsilon)].
\label{eq:root-vt}
\end{eqnarray}

\subsection{FCT-SHASTA Algorithm}

The conservation equations in Eq.~(\ref{eq:hydro2}) have the general form of coupled convective diffusion equations which can be solved using 
extended FCT (Flux-Corrected Transport)-SHASTA algorithm \cite{Boris:1973cp, Zalesak:1979, Rischke:1995ir}. Here we give a brief 
overview of the algorithm using the following 1D partial differential equation as an example:
\begin{equation}
  \partial_{t} \rho + \partial_{x} (v \rho) = 0 .
  \label{eq:shasta_demo}
\end{equation}

The FCT-SHASTA algorithm first evolves this equation by a transport and diffusion stage which ensures the solution's 
monotonicity and positivity,
 \begin{eqnarray}
	 \rho_{j}^{td}  &=&  \frac{1}{2}Q_{-}^{2}(\rho_{j-1}^{n}-\rho_{j}^{n})  +  \frac{1}{2}Q_{+}^{2}(\rho_{j+1}^{n}-\rho_{j}^{n})
	\nonumber  \\
	               & & +(Q_{+}+Q_{-})\rho_{j}^{n},\\	
Q_{+}&=&(\frac{1}{2}-v_{j}^{1/2}\frac{\delta t}{\delta x})/\left[1+(v_{j+1}^{1/2}-v_{j}^{1/2})\frac{\delta t}{\delta x}\right],\\
Q_{-}&=&(\frac{1}{2}+v_{j}^{1/2}\frac{\delta t}{\delta x})/\left[1-(v_{j-1}^{1/2}-v_{j}^{1/2})\frac{\delta t}{\delta x}\right],
	   \label{eqn:QpQm}
	 \end{eqnarray}
where $td$ stands for ``transport and diffusion'', $j$ denotes the discretized space index in $\delta x$ and $n$ 
the time step in $\delta t$, $v_{j}^{1/2}$ denotes the value of $v_{j}$ at half time-step  $n+1/2$ which 
is evaluated with a 2-step Runge-Kutta method.
The derivation of $Q_{-}$ and $Q_{+}$ can be found in the Appendix.
In the zero-velocity limit (where $Q_{+} = Q_{-} = 1/2$) the above solution becomes,
\begin{eqnarray}
  \rho_{j}^{td} &=& \rho_{j}^{n} - \frac{1}{8} (\rho_{j+1}^{n} -2\rho_{j}^{n}+\rho_{j-1}^{n}) \nonumber \\
  &=& \rho_{j}^{n} - f_{j+1/2} + f_{j-1/2},
  \label{eq:diff_flux}
\end{eqnarray}
from which one can clearly identify the diffusion term  $\frac{1}{8}(\rho_{j+1}^{n} - 2\rho_{j}^{n} + \rho_{j-1}^{n})$. 
The general form of the diffusion can be expressed by the flux $f_{j\pm 1/2} = \pm \frac{1}{8}(\rho_{j\pm 1}-\rho_{j})$.
In the second stage of the FCT-SHASTA algorithm, an anti-diffusion term is added to the transported and diffused result.
Usually the anti-diffusion term are calculated in the FCT by subtracting low order from high order 
transport results to make sure the anti-diffusion is accurate enough.
Since ripples arise in high order transport algorithm, the flux in the anti-diffusion term must be 
corrected to make sure no new maximum or minimum are produced.
In our calculation, the flux limiter developed by Zalesak \cite{Zalesak:1979} in the anti-diffusion stage
is extended to a full multi-dimensional FCT algorithm.  In this multi-dimensional algorithm the
1D FCT-SHASTA algorithm with time-splitting \cite{Rischke:1995ir} is used along one direction at a split-time step, while a $x\rightarrow y \rightarrow \eta_{s} \rightarrow y\rightarrow x$ rotation  is used to extend the FCT-SHASTA algorithm to multi-dimensions and to suppress the numerical eccentricity produced in transverse direction during the hydrodynamic evolution.



We will use the second order midpoint Runge-Kutta method to include the source term $S$ in a differential equation,
\begin{equation}
  \frac{d \rho }{d t} = S,
  \label{}
\end{equation}
which involves two steps,
\begin{enumerate}
  \item $\rho^{n+1/2} = \rho^{n} + \frac{1}{2} \delta t S(t, \rho^{n})$
  \item $\rho^{n+1}   = \rho^{n} + \delta t S(t+\delta t/2, \rho^{n+1/2})$ 
\end{enumerate}
The energy density and velocity calculated in half time step $\delta t/2$ can be used in the FCT-SHASTA algorithm to improve the numerical precision as described in the Appendix. The 2nd order Runge Kutta method can significantly improve the numerical accuracy and remove the numerical diffusion for much larger time steps.

We use the following simplified conservation equation to describe the numerical method and procedures of a 
combined FCT-SHASTA and the 2nd order Runge-Kutta algorithm in solving hydrodynamic equations in our study,
\begin{equation}
  \partial_{\tau} {\cal T} + \partial_{i}(v_{i} {\cal T} ) = S,
  \label{eq:demo}
\end{equation}
where we use ${\cal T} \equiv \tau T^{\tau\nu}$ to denote one component of the energy-momentum tensor.

\begin{enumerate}
  \item Calculate source term at time step $n$: $S=S(\tau^{n}, \varepsilon^{n}, v_{i}^{n}, {\cal T}^{n})$.
  \item Evolve $\partial_{\tau}{\cal T} + \partial_{i}(v_{i}{\cal T}) = 0$ to time step $n+1/2$ by using SHASTA algorithm to get ${\cal T}^{\prime n+1/2}$.
  \item Update to ${\cal T}^{n+1/2} = {\cal T}^{\prime n+1/2} + 0.5\Delta \tau S(\tau^{n}, \varepsilon^{n}, v_{i}^{n}, {\cal T}^{n})$ and use the root-finding method to calculate the energy density and velocity $\varepsilon^{n+1/2}, v_{i}^{n+1/2}$.
  \item Calculate source term at half-time step $n+1/2$: $S=S(\tau^{n+1/2}, \varepsilon^{n+1/2}, v_{i}^{n+1/2}, {\cal T}^{n+1/2})$.
  \item Evolve $\partial_{\tau}{\cal T} + \partial_{i}(v_{i}{\cal T}) = 0$ to time step $n+1$ by using SHASTA algorithm with half time step velocity $v_{i}^{n+1/2}$ to obtain ${\cal T}^{\prime n+1}$.
  \item Update to ${\cal T}^{n+1} = {\cal T}^{\prime n+1} + \Delta \tau S(\tau^{n+1/2}, \varepsilon^{n+1/2}, v_{i}^{n+1/2}, {\cal T}^{n+1/2})$ and calculate the energy density and velocity for the next time step  $\varepsilon^{n+1}, v_{i}^{n+1}$ via root-finding method.

\end{enumerate}
We refer readers to the Appendix for more details about the SHASTA algorithm. We have used $1D$ SHASTA algorithm 
with time-splitting \cite{Rischke:1995ir} to solve the hydrodynamic equations and it is easy to implement parallel computing in the future. In the current event-by-event simulations, only high level parallel computing is used where events run on separate CPU's. The most time consuming part in hydrodynamic simulations is to calculate spectra of direct thermal hadrons and decay products from hundreds of resonances for comparison with experimental data. The CPU hours used in our simulations are significantly reduced by our improved algorithm 
to calculate the freeze-out hyper-surface.

\subsection{Freeze Out and Hadronization}

We will use the Cooper-Frye formula \cite{Cooper:1974mv}  to calculate the momentum distribution for 
particle $i$ with degeneracy $g_{i}$:
\begin{equation}
  \label{eq:frye-cooper}
  E\frac{dN_{i}}{d^{3}P}=\frac{dN_{i}}{dYp_{T}dp_{T}d\phi}=g_{i}\int_{\Sigma}p^{\mu}d\Sigma_{\mu}f(p\cdot u),
\end{equation}
where $d\Sigma_{\mu}$ is the normal vector of a small piece of freeze-out hyper-surface beyond which the 
temperature falls below the freeze-out temperature $T_{f}$ or energy density $\varepsilon$ falls 
below the freeze-out density $\varepsilon_f$. Hadrons pass through the freeze-out surface element 
is assumed to obey thermal distribution at temperature $T_{f}$,
\begin{equation}
  f(p\cdot u)=\frac{1}{(2\pi)^{3}}\frac{1}{e^{((p\cdot u - \mu_{i})/T_{f}))}\pm 1},
  \label{eq:fermion-bosen}
\end{equation}
where $\pm$ stands for fermions and bosons respectively, $u$ is the flow velocity.
All resonances are assumed to freeze out from the same hyper surface and decay into stable particles.
The invariant energy of particle in comoving frame is,
\begin{eqnarray}
   E&=& p\cdot u  \\
   &=&  u^{\tau}\left[ m_T \cosh(Y-\eta_s) - \vec p_\perp\cdot \vec v_\perp 
   - m_T \sinh(Y-\eta_s)v_{\eta}\right]\nonumber
  \label{eq:inv_E}
\end{eqnarray}
where $p^{\mu} = (m_T \cosh(Y-\eta_s) , \vec p_\perp  , m_T \sinh(Y-\eta_s)/\tau )$.
In order to calculate the spectra, we need to know the freeze-out hyper surface $\Sigma = (\tau_f, x, y, \eta_s)$ at freeze-out time $\tau_f$. The normal vector for one piece of freeze-out hyper surface in invariant-time coordinate is, 

\begin{equation}
  d \Sigma_\mu =  ( \tau_f dx dy d\eta_s, -\tau_f d\tau dy d\eta_s, -\tau_f d\tau dx d\eta_s, -d\tau dx dy )
  \label{eq:frzsf}
\end{equation}

In a simple cuboidal method by Hirano \cite{Hirano:2001eu}, finite grid 
sizes $\Delta\tau$, $\Delta x$, $\Delta y$ and $\Delta \eta_s$ are used to calculate $d\Sigma_{\mu}$.
The surface elements are calculated independently for each direction. 
In the $\tau$-direction, a cuboidal volume $d\Sigma_{\tau} = \tau_f \Delta x \Delta y \Delta \eta_s$ is recorded when the freeze-out temperature falls between $T(\tau_{n}, x,y,\eta_s)$ at time step $n$ and $T(\tau_{n+1},x,y,\eta_s)$ at time step $n+1$. The norm vector points to the low energy density direction along the $\tau$ axis. The freeze-out time $\tau_f$ and 4-velocity $u$ are calculated from interpolation between time step $\tau_{n}$ 
and $\tau_{n+1}$.
In the $x$-direction, $d\Sigma_{x} = - \tau \Delta \tau \Delta y \Delta\eta_s$ is recorded at time step $n$ where $T_f$ falls between $ T(\tau_{n}, x_i, y, \eta_s)$  and $T(\tau_{n}, x_{i+1}, y, \eta_s)$. 
Hirano's method seems to overestimate the freeze-out hyper surface in one single cell where the total cuboidal volume is always added without considering the cut-through position. After the decomposition of $p^{\mu}d\Sigma_{\mu}$ to $4$ directions across several hydrodynamic cells, one finds that the overestimated part in one cell actually fills up the underestimated part in another.  It is quite a good approximation as long as the velocity at the freeze-out hyper surface does not change too much and its $3$ components can be treated as the flow velocity on the cube edge. The requirement can be easily fulfilled by using one single hydro cell cube.
The problem is that the size of the data file for the whole freeze-out hyper surface strongly depends on the time and space grid size used
in solving the hydrodynamic equations. If a smaller grid size is used (for example $\Delta \tau = 0.01$ fm and $\Delta x = 0.1$ fm) 
to improve the numerical accuracy in the transport stage, the data file becomes huge and the calculation of hadron spectra for hundreds of resonances will be very time consuming.

To improve the computation efficiency for finer grids, one can divide the whole hyper surface into smaller pieces inside interpolation 
cubes each extending to several hydrodynamic grid cells along $4$ directions \cite{Kataja:1990tp, Sollfrank:1996hd, Kolb:2000sd, Schenke:2010nt}. 
Each piece of surface elements is presented by the intersections $s_{i}$  on edges of an interpolation cube where the hyper surface cuts through, and its area is approximated by a group of triangles (in the (2+1)D case) or a group of tetrahedra (in the (3+1)D case) formed by these intersections. The main task is to triangulate these intersections in (2+1)D or (3+1)D hydro, and calculate the areas of the triangles or volumes of the tetrahedra piece by piece.  

\begin{figure}[t]
  \begin{center}
	\includegraphics[scale=0.35]{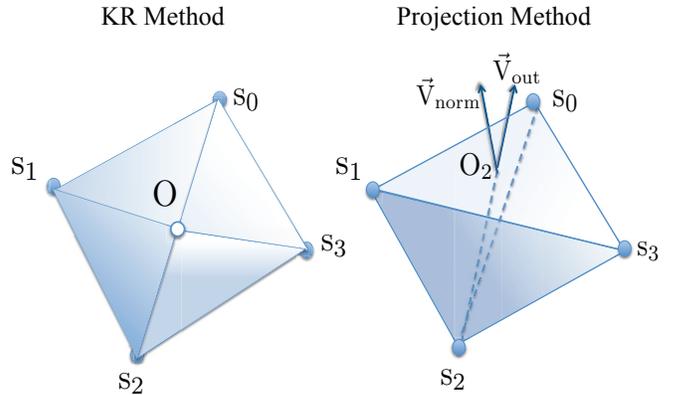}
  \end{center}
  \caption{(Color online) An example of (2+1)D freeze-out hyper surface. In the projection method, 
  there are $4$ triangles formed on the convex hull $\Diamond s_{0}s_{1}s_{2}s_{3}$. Only two are 
  chosen by the criterion $\vec{V}_{norm}\cdot \vec{V}_{LE}>0$, where $\vec{V}_{LE}$ is the low 
  energy density direction and $\vec{V}_{norm}$ is the outward normal vector of each surface triangle.}
  \label{fig:frz_hpsf}
\end{figure}

In the algorithm developed by Kataja, Ruuskanen (KR) and collaborators \cite{Kataja:1990tp,Sollfrank:1996hd} and later used 
in the Azhydro code by Kolb \cite{Kolb:2000sd} for the (2+1)D case, 
these intersections on the edges of an interpolation cube are ordered into a circular sequence. The area $S$ of one piece of hyper surface inside 
an interpolation cube is approximated by the summation of the areas of a group of triangles with each triangle constructed by connecting 
two nearby intersections with the center point $O$ of all the intersections 
(as shown in the left panel of Fig.~\ref{fig:frz_hpsf}),
\begin{equation}
  S=\sum_{i=1}^{N}\triangle Os_{i}s_{i+1},
  \label{eq:azhydro-hpsf}
\end{equation}
where $N$ is the number of intersections and $s_{N+1}=s_{1}$. The flow velocity and energy density at the center 
point $O$ for this piece of surface element used in the Cooper-Frye formula is approximated by averaging over all the 
intersection points. In the KR method for the (2+1)D case, the most difficult part is to order these intersections in a circular 
sequence and this is achieved by using a bit-chart of the distances between the mid-points of any 2 of the 12 edges of
the interpolation cube. Once the intersections are ordered, the area of one triangle formed by the 
center point $O$ and two neighboring intersections can be calculated from:
\begin{equation}
  \vec{V}_{norm}^{2+1D} =\frac{1}{2} 
  \begin{vmatrix}
	n & i & j \\ A_{0} & A_{1} & A_{2} \\ B_{0} & B_{1} & B_{2}
  \end{vmatrix}
  = n d\Sigma_{0}+i d\Sigma_{1}+j d\Sigma_{2},
  \label{eq:Vnorm2}
\end{equation}
where $A$ and $B$ are the two vectors that span the triangle in (2+1)D hydro.

The above KR method has been extended to (3+1)D hydro  recently by the McGill group \cite{Schenke:2010nt}. 
A bit-chart for 32 edges on the $d\tau dx dy d\eta_s$ super cube was constructed. The triangles are replaced by tetrahedra.
The volume of one tetrahadron is then,
\begin{equation}
  \vec{V}_{norm}^{3+1D} = \frac{1}{6}
  \begin{vmatrix}
	n & i & j & k\\ A_{0} & A_{1} & A_{2} & A_{3} \\ B_{0} & B_{1} & B_{2} & B_{3} \\ C_{0} & C_{1} & C_{2} & C_{3} 
  \end{vmatrix}
  =n d\Sigma_{0}+i d\Sigma_{1}+j d\Sigma_{2}+k d\Sigma_{3},
  \label{eq:Vnorm3}
\end{equation}
where $A$, $B$ and $C$ are the three vectors that span the tetrahedron in the (3+1)D case.

In this paper, we develop a projection method to calculate the freeze-out hyper surface $d\Sigma_{\mu}$ in our (3+1)D 
hydrodynamic simulation. We illustrate the method here for the (2+1)D case as shown in the right panel of Fig.~\ref{fig:frz_hpsf}.
The extension to (3+1)D or (n+1)D is straightforward. Here we list a step-by-step procedure for calculating the freeze-out
hyper surface in our projection method:
\begin{itemize}
\item Identify interaction points whenever energy density difference $\epsilon -\epsilon_{\rm fr}$ changes sign between two grid points, where $\epsilon_{\rm fr}$ is
the freeze-out energy density.
 
  \item An ``n-simplex'' is an n-dimensional polytope which is the convex hull of its n+1 vertices. A 2-simplex is a triangle, a 3-simplex is a tetrahedron, a 4-simplex is a pentachoron \cite{Wiki:n-simplex}.
  
  \item One can calculate the area of one piece of hyper surface directly if there are only $3$ intersections on the edges of 
  the interpolation cube. For $4$ or more intersections,  select any 4 intersections $s_{i=0,1,2,3}$ to construct 
  one 3-simplex or tetrahedron $\Diamond s_{0}s_{1}s_{2}s_{3}$ that has 4 triangles on its surface 
  as shown in the right panel of Fig.~\ref{fig:frz_hpsf}.

  \item Consider one triangle  $\triangle s_{0}s_{1}s_{3}$ whose center is denoted by $O_{2}$. The vector between the fourth
  intersection point and the center of the chosen triangle, $\vec{V}_{out}=\vec{s_{2}O_{2}}$ is defined as the outward direction of the tetrahedron on $\triangle s_{0}s_{1}s_{3}$ side. In the (3+1)D case, $\vec V_{out}$ is defined as $\vec{s_{n}O_{n}}$ where $s_{n}$
 is the intersection point opposite to center of the tetrahedron $O_{n}$.   
  
  \item The normal vector of $\triangle s_{0}s_{1}s_{3}$ has two directions (this is also true for 3+1D case).  Choose the
  outward normal vector $\vec{V}_{norm}$ that satisfies $\vec{V}_{norm}\cdot \vec{V}_{out}>0$.
    
 \item For an interpolation cube with more than 4 intersections, select another intersection $s_{i}$ outside 
 the first constructed 3-simplex or tetrahedron $\Diamond s_{0}s_{1}s_{2}s_{3}$.  This
 intersection $s_{i}$ is considered out of the tetrahedron on a triangle $\triangle s_{0}s_{1}s_{3}$ side
  when $\vec{V}_{norm}\cdot \vec{O_{2}s_{i}}>0$. Find all possible triangles of the tetrahedron that the intersection
  $s_{i}$ is out of. Form a new tetrahedron between $s_{i}$ and each of these triangles. Remove those triangles
  that are shared between the first and the new tetrahedra to form a new convex hull.
  
  \item Repeat the above step for the rest of intersections and the previous formed convex hull until a closed convex hull 
  is formed with all the triangles that we are interested in.
  
   \item Consider $\vec v_{i LE}$ as the low energy density flow vector for each intersection point on one
  triangle $\triangle s_{0}s_{1}s_{3}$. Then define $\vec V_{LE}=\sum_{i=0,1,3} \vec v_{i LE}$ as the low energy density flow
  vector of the triangle $\triangle s_{0}s_{1}s_{3}$. This triangle will be considered as a piece of the freeze-out hyper surface
  only when $\vec{V}_{norm}\cdot \vec{V}_{LE}>0$.

  \item Use the criteria $\vec{V}_{norm}\cdot \vec{V}_{LE}>0$ to select all the triangles of the closed convex hull that will form
  the freeze-out hyper surface of the interpolation cube, with contribution from each triangle given by $\vec{V}_{norm}$.
  
\end{itemize}
The above algorithm can be easily extended to the (3+1)D case with 3-simplex replaced by 4-simplex and triangles
replaced by tetrahedra whose $\vec{V}_{out}$, $\vec{V}_{norm}$ and $\vec{V}_{LE}$ can be similarly defined. 

If $4$ intersections are coplanar in (2+1)D or $5$ intersections are co-tetrahedron in (3+1)D hydro, random tiny placements
with an amplitude of $10^{-9}$ will be applied to the intersections before the above procedure is applied in our algorithm.
 A numerical error of the order of $10^{-9}$  for the hyper surface calculation is negligible but this will keep 
 the algorithm continue to run for all possible hyper surfaces. This has been checked for calculating the volume 
 of a 3D cube with the same time coordinate $\tau$ in $4$ dimensional space.

In all above methods, if the hyper surface cuts through the same edge of the interpolation cube more than once but odd
times, the intersections are approximated by a single point. An even number of intersections on a single edge of the 
interpolation cube are neglected. This approximation can be improved in the future by refining our projection method
at the expense of minimal increase of computer time. We have compared our projection method with Hirano's cuboidal
freeze-out method, and find very tiny discrepancies in multiplicity distribution and $p_{T}$ spectra. 
However, for small grid size, our projection method is much faster in calculating the final particles' spectra, 
especially in event-by-event simulations of higher order harmonic flows.

\subsection{Resonance decays}

To calculate the final hadron spectra from hydrodynamic simulations, one needs to include both direct thermal hadrons
that go through the freeze-out surface and decay products from resonances which are assumed to
freeze out at the same temperature. Resonances with mass up to $1.68$ GeV are considered in the 
current calculation. 

We have extended the numerical procedure for 
resonance decay from Kolb's (2+1)D Azhydro code to our (3+1)D hydrodynamic simulations in this study. 
In the original Azhydro code, Bjorken scaling is assumed for the rapidity distributions of thermal hadrons 
and resonances. Final hadron spectra at any given rapidity will contain decay products from resonances
in all rapidities with a uniform and infinitely long distribution.  In our (3+1)D hydro calculation, this uniform
and infinite rapidity distribution is replaced with the more realistic one that are not smooth in each event as
determined by the initial
condition from the AMPT model. In our numerical calculations, we still have to limit the rapidity range
to a finite value $[-8,8]$ and assume a linear interpolation outside
this range assuming yields for all resonances to be zero at $Y =\pm 20$.

\subsection{Initial Conditions}

To incorporate fluctuations and correlations in both transverse and longitudinal flow velocities in event-by-event
(3+1)D hydrodynamic simulations, we will use the AMPT model \cite{Zhang:1999bd} to provide the local initial energy-momentum 
tensor in each hydrodynamic cell. The AMPT model uses the HIJING model \cite{Wang:1991hta, Gyulassy:1994ew,Wang:2009qb}
to generate initial partons from hard and semi-hard scatterings and excited strings from soft interactions.  
The number of excited strings in each event is equal to that of participant nucleons.  The number of
mini-jets per binary nucleon-nucleon collision follows a Poisson distribution with the average number given by the
mini-jet cross section, which depends both on the colliding energy and the impact parameter through an impact-parameter 
dependent parton shadowing \cite{Wang:1991hta} in a nucleus.  In this model, the total local energy-momentum 
density of partons and its fluctuations will be determined by the number of participants, binary
nucleon-nucleon collisions, number of mini-jets per nucleon-nucleon collision and the fragmentation
of excited strings. HIJING uses the Glauber model to determine the number of participants and binary nucleon-nucleon 
collisions with the Wood-Saxon nuclear distribution. 

The formation times for partons from mini-jets produced via semi-hard scatterings are short. Their energy-momentum density 
can be used as part of the initial conditions for the hydrodynamic evolution. However, strong color fields in the soft strings
take time to materialize and their contribution to the initial energy-momentum density at earlier times is hard to estimate.
In the option that we use in AMPT, strings are melt via conversion of hadrons into quarks and 
anti-quarks after the string fragmentation which will participate in the parton cascade together with hard and semi-hard partons.
The formation time of these soft partons are estimated according to their transverse momentum 
and energy ($ t_{f} \sim 2p_{0}/p_{T}^{2}$). In our study we allow AMPT model to run through 
the parton cascade for an initial period of time. We record the space-time points of the last scattering
or formation time for all the partons. Most of the partons are found to concentrate along the hyperbola of an initial proper time $\tau_{0}$.
As an approximation, we simply assign the proper time $\tau_{0}$ to all partons and use their 4-momenta to calculate the local
energy-momentum tensor as the initial condition for our ideal hydro evolution. In this approximation for the initial conditions at a given proper time,  parton interaction at large spatial rapidity at very late Cartesian time in the AMPT model is neglected. 
These are questionable approximations that one has to keep in mind when one considers theoretical uncertainties and future improvements. In principle, one should run the AMPT model to the end (no further interactions). The recorded particles' (both partons and hadrons) space-time positions when they cross the hyperbola with fixed $\tau_{0}$ will provide the initial condition for the hydrodynamical evolution. This is, however, too demanding in computer time.

The $4$-momenta and space coordinates of partons from the AMPT model according to the above description will be used 
to calculate the local energy-momentum tensor as the initial conditions for our 
event-by-event (3+1)D hydrodynamic simulations. Its value in each grid cell is approximated by a gaussian 
distribution in invariant-time coordinates,
\begin{equation}
  \begin{aligned}
  T^{\mu\nu} &(\tau_{0},x,y,\eta_{s}) = K\sum_{i}
  \frac{p^{\mu}_{i}p^{\nu}_{i}}{p^{\tau}_{i}}\frac{1}{\tau_{0}\sqrt{2\pi\sigma_{\eta_{s}}^{2}}}\frac{1}{2\pi\sigma_{r}^{2}}\\
     	  &\hspace{-0.1in} \times \exp \left[-\frac{(x-x_{i})^{2}+(y-y_{i})^{2}}{2\sigma_{r}^{2}} - \frac{(\eta_{s}-\eta_{i s})^{2}}{2\sigma_{\eta_{s}}^{2}}\right],
  \end{aligned}
  \label{eq:Pmu}
\end{equation}
where $p^{\tau}_{i}=m_{iT}\cosh(Y_{i}-\eta_{i s})$, $p^{x}_{i}=p_{i x}$, $p^{y}_{i}=p_{i y}$ 
and $p^{\eta}_{i}=m_{i T}\sinh(Y_{i}-\eta_{i s})/\tau_{0}$ for parton $i$, which runs over all partons produced in the AMPT 
model simulations. 
We have chosen $\sigma_{r}=0.6$ fm, $\sigma_{\eta_{s}}=0.6$ in our calculations.
The transverse mass $m_{T}$, 
rapidity $Y$ and  spatial rapidity $\eta_{s}$ are calculated from the parton's $4$-momenta and spatial coordinates. Note here 
that the Bjorken scaling assumption $Y=\eta_{s}$ is not used here because of early parton cascade before the initial time and the uncertainty
principle applied to the initial formation time in AMPT. The scale factor $K$ and the initial time $\tau_{0}$ 
are the only two parameters that one can adjust to fit the experimental data on central rapidity density of produced hadrons. 


Note that the Gaussian smearing in  Eq.~(\ref{eq:Pmu}) smoothes out the energy-momentum tensor within several hydro grid
cells. Such a smearing acts like an initial thermalization process similar to the parton cascade in the AMPT model within
the initial time $\tau_{0}$. The initial matter in each grid cell is then assumed to reach a local thermal equilibrium and
one can obtain the initial local energy density and flow velocity from Eq.~(\ref{eq:Pmu}) 
using the root finding method in Eqs.~(\ref{eq:root_finding}-\ref{eq:root-vt}). This is equivalent to the prescription in Ref.~\cite{Gardim:2011qn}
for a constant proper time surface.

\section{Hadron spectra from event-by-event hydrodynamics}

In this section, we will compare hadron spectra from our event-by-event (3+1)D ideal hydrodynamic simulations
to experimental data at RHIC and LHC energies. For $Au+Au$ collisions at the RHIC energy $\sqrt{s}=200$ GeV,
we use a scale factor $K=1.45$ and an initial time $\tau_{0}=0.4$ fm/$c$ in the initial conditions from the AMPT model.
We have used grid spacings $\delta x=\delta y=0.3$ fm, $\delta \eta_{s}=0.2$ and $\delta \tau=0.04$ fm/$c$ with
grid size $L_{x}\times L_{y}\times L_{\eta}=(30 fm)\times(30 fm)\times 6$.
For $Pb+Pb$ collisions at the LHC energy $\sqrt{s}=2.76$ TeV, we use $K=1.6$, $\tau_{0}=0.2$ fm/$c$, 
 $\delta x=\delta y=0.2$ fm, $\delta \eta_{s}=0.3 $ and $\delta \tau=0.03$ fm/$c$ and 
grid size $L_{x}\times L_{y}\times L_{\eta}=(30 fm)\times(30 fm)\times 12$. With these grid spacings and
sizes, we have checked that the increase of the total entropy of the system due to the numerical viscosity 
over the entire evolution duration of about 20 fm/$c$ is less than 1\%.

\begin{figure}[h]
  \begin{center}
	\includegraphics[scale=0.45]{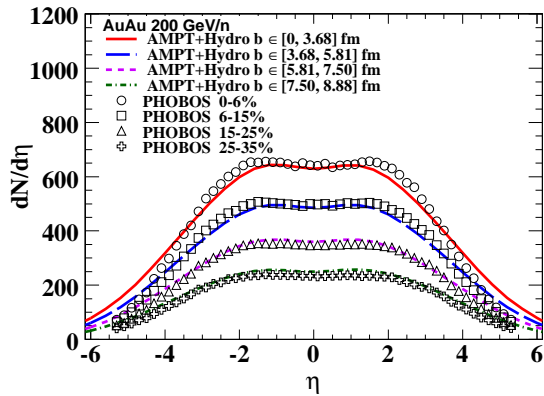}
  \end{center}
  \caption{(Color online) Charged hadron pseudo-rapidity distributions from event-by-event (3+1)D ideal hydro (lines)
  with AMPT fluctuating initial conditions compared to the PHOBOS experimental data \cite{nucl-ex/0210015} (symbols)
  for $Au+Au$ collisions at the RHIC energy $\sqrt{s}_{NN}=200$ GeV. Centralities and the corresponding
  ranges of impact parameters are given by STAR's Glauber model results \cite{STAR:2008ez}.}
  \label{fig:dNdEta}
\end{figure}

\begin{figure}[h]
  \begin{center}
    \includegraphics[scale=0.45]{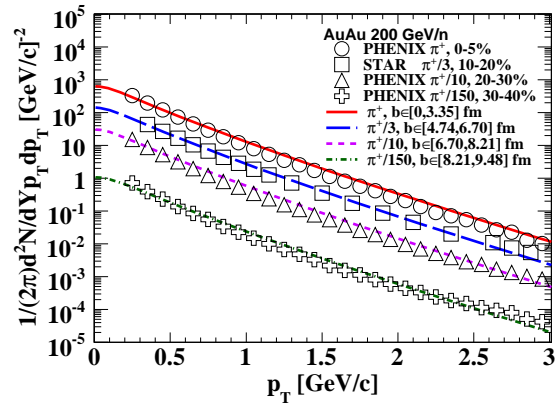}
  \end{center}
  \caption{(Color online)  Transverse momentum spectra for $\pi^{+}$ from event-by-event (3+1)D ideal hydro (lines)
  with AMPT fluctuating  initial conditions compared to the PHENIX \cite{ADLER:2003CB} and 
  STAR experimental data \cite{nucl-ex/0606003} (symbols) for  $Au+Au$ collisions at different centralities 
  at the RHIC energy $\sqrt{s}_{NN}=200$ GeV}
       \label{fig:dNdPt}
\end{figure}

\begin{figure}[h]
  \begin{center}
	\includegraphics[scale=0.45]{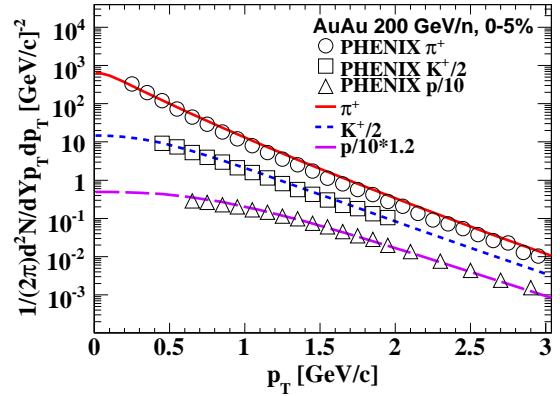}
  \end{center}
  \caption{(Color online) Transverse momentum spectra for identified particles from event-by-event (3+1)D ideal 
  hydro (lines)  with AMPT fluctuating  initial conditions compared to the PHENIX experimental data (symbols) 
  \cite{ADLER:2003CB} for the most central (0-5\%) $Au+Au$ collisions at the RHIC energy $\sqrt{s}_{NN}=200$ GeV.
  A factor of $1.2$ is multiplied to the hydro proton spectra due to possible early chemical freeze-out.}

  \label{fig:dNdPt_identified}
\end{figure}

Shown in Figs.~\ref{fig:dNdEta}, \ref{fig:dNdPt} and \ref{fig:dNdPt_identified} are
pseudo-rapidity distributions for charged hadrons, $p_{T}$ spectra for charged pions 
and $p_{T}$ spectra for identified charged hadrons, respectively, from our event-by-event 
ideal hydrodynamic calculations. Also shown are experimental data for $Au+Au$ collisions
with different centralities at the RHIC energy.

The experimental data on central pseudo-rapidity density of charged hadrons in the most central $0-5\%$ 
 collisions are used to calibrate the parameters, $K=1.45$ and $\tau_{0}=0.4$ fm/$c$, in our model for the initial conditions. 
 For other centralities, the range of impact parameters is varied according to the Glauber model \cite{STAR:2008ez}. 
We assumed the freeze-out temperature $T_{f}=0.137$ GeV and the parameterized equation of 
state EoSL s95p-v1 \cite{Huovinen:2009yb}. The spectra for charged hadrons include both direct thermal 
hadrons and decay products from resonances with masses up to $1.68$ GeV. All hadron spectra agree with 
the experimental data well for all centralities and $p_{T}<3$ GeV/$c$ except for protons
which are about a factor of 1.2 smaller than the data. The deficiency in proton spectra might be caused by
finite chemical potential due to chemical freeze-out time earlier than the pions and kaons \cite{Kolb:2003dz, Huovinen:2011xc}.
The charged hadron yields at large rapidity from our (3+1)D hydro calculation (Figs.~\ref{fig:dNdEta}) are somewhat 
larger than the experimental data. In these large rapidity regions, the net baryon density is quite large. One has to 
consider the evolution of the net baryon density coupled to the energy-momentum density and the
EoS we used for zero baryon chemical potential is no longer valid. Inclusion of shear viscosity also
slows down the longitudinal expansion and gives a narrower tail of the pseudo-rapidity distribution of
charged hadrons \cite{Schenke:2011bn,Bozek:2011ua}.

The overall geometric shape of the overlapped region and local density fluctuation of the
initially produced dense matter will lead to azimuthal anisotropies in the final hadron spectra. 
The strong elliptic flow measured during the first years of RHIC experiments is considered as one of
the evidences for a strongly coupled quark-gluon plasma (sQGP) formed in the central $Au+Au$ 
collisions \cite{Gyulassy:2004zy}. Recent efforts have been focused on extracting the shear viscosity of
the sQGP by comparing experimental data with viscous hydrodynamic 
calculations \cite{Romatschke:2007mq, Song:2007ux,Dusling:2007gi,Song:2010mg}. However, quantitative studies are
complicated by uncertainties in the initial conditions \cite{Song:2011qa}. Inclusion of fluctuations will add to the complexities
of the problem. It is therefore useful to study the variation of anisotropic flows with more realistic initial conditions
even in ideal hydrodynamics without viscous corrections.
Most recent studies \cite{Petersen:2010cw,Holopainen:2010gz,Schenke:2010rr,Albacete:2011fw, Schenke:2011bn,  Qiu:2011hf, Schenke:2012wb,Gardim:2012yp} employed either Monte Carlo Glauber\cite{Miller:2007ri}  or Monte Carlo  KLN \cite{Drescher:2006ca} 
initial conditions, both lack fluctuations in local flow velocity and longitudinal distribution in pseudo-rapidity.

The differential harmonic flow $v_{n}$ of hadron spectra is defined as:
\begin{equation}
  v_{n}(p_{T},\eta)=\frac{\int_{0}^{2\pi}\,d\phi \frac{dN}{d\eta p_{T}dp_{T}d\phi}\cos\bigl(n(\phi-\Psi_{n})\bigr)}{\int_{0}^{2\pi}\,d\phi \frac{dN}{d\eta p_{T}dp_{T}d\phi}},
  \label{eq:vn_pt}
\end{equation}
where $\Psi_{n}$ can be the azimuthal angle for the participant-plane (PP) in coordinate space of initial partons in theoretical studies
or event-plane (EP)  in momentum space of the final hadrons in experimental analyses,
\begin{subequations}
  \label{eqn:EP_PP}
  \begin{eqnarray}
	\Psi_{n}^{\rm PP}&=& \frac{1}{n}\bigl(\arctan \frac{\langle r^{2}\sin(n\phi_{r}) \rangle}{\langle r^{2}\cos(n\phi_{r}) \rangle} +\pi \bigr) ,
	\label{equationa}\\ 
	\Psi_{n}^{\rm EP}&=& \frac{1}{n}\arctan \frac{\langle p_{T}\sin(n\phi_{p}) \rangle}{\langle p_{T}\cos(n\phi_{p}) \rangle}.
	\label{equationb} 
  \end{eqnarray}
\end{subequations}
The average in Eq.~(\ref{equationa}) for $\Psi_{n}^{\rm PP}$ is over all initial partons weighted by their
squared transverse coordinates $r^{2}=x^{2}+y^{2}$, 
while the average in Eq.~(\ref{equationb}) for $\Psi_{n}^{\rm EP}$ is over final particles 
weighted by their transverse momenta. The corresponding hadronic flows will be denoted as $v_{n}^{\rm PP}$ and $v_{n}^{\rm EP}$,
respectively. Note that the final hadron spectrum from Coorper-Frye formula 
is a continuous distribution function. Therefore, integrations over the transverse momentum $p_{T}$, pseudo-rapidity $\eta$ 
and azimuthal angle $\phi_{p}$ in calculating $\Psi_{n}^{\rm EP}$ will not introduce plane resolution due to finite number of 
particles per event which will have to be corrected for in experimental analyses.

\begin{figure}[t]
  \begin{center}
	\includegraphics[scale=0.45]{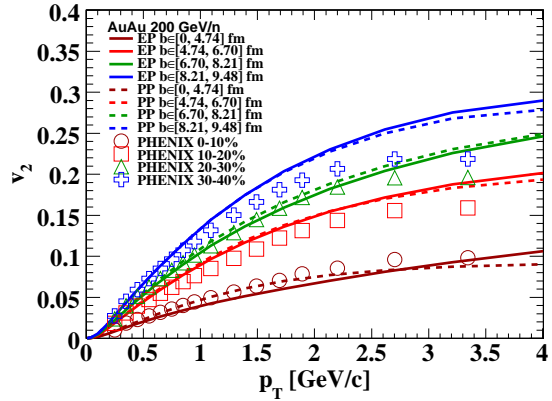}
  \end{center}
  \caption{(Color online) Elliptic flow for charged hadrons with respect to participant-planes (PP) (solid) and event-planes (EP) (dashed)
  from event-by-event (3+1)D hydrodynamic simulations with AMPT initial conditions compared to the PHENIX data \cite{Adare:2011tg} (symbols)  on $v_{2}^{\rm EP}$ for $Au+Au$ collisions at the RHIC energy $\sqrt{s}_{NN}=200$ GeV.}
  \label{fig:v2_EP_PP}
\end{figure}

In Fig.~\ref{fig:v2_EP_PP}, we compare our ideal hydro calculation of $v_{2}^{\rm EP}$  
(solid lines) with the PHENIX data \cite{Adare:2011tg} for charged hadrons within a pseudo-rapidity 
range $[-0.35,0.35]$ in $Au+Au$ collisions at $\sqrt{s}_{NN}=200$ GeV with different centralities. The event-planes
in the PHENIX analysis were determined with $2$ sub-events to correct for the event-plane resolution.
Our hydro calculations fit the
experimental results quite well at low $p_{T}$ for all centralities. At higher $p_{T}$, viscous corrections
and other non-equilibrium effects such as jet quenching \cite{Wang:2000fq,Gyulassy:2000gk,Bass:2008rv}  
are expected to become important. Ideal hydrodynamics will fail, producing much larger elliptic flow than 
the experimental data. We also show $v_{n}^{\rm PP}$
(dashed lines) from our hydro calculations as determined by the participant-plane. It is a very good approximation
of $v_{n}^{\rm EP}$ as determined by the event-plane, especially at low $p_{T}$. In the rest of this paper,
we will focus on the elliptic flow $v_{2}^{\rm EP}$ with respect to event-planes.

\begin{figure}[t]
  \begin{center}
	\includegraphics[scale=0.45]{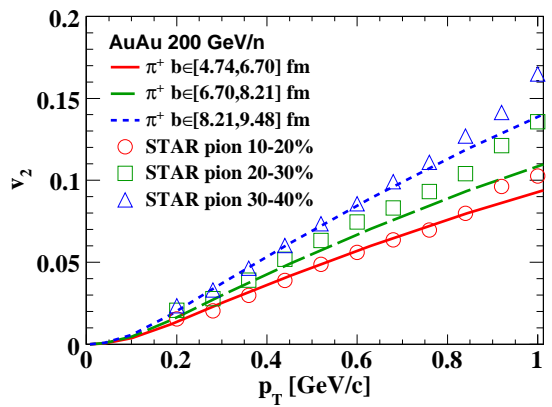}
  \end{center}
  \caption{(Color online) Elliptic flow for $\pi^{+}$ at different centralities from event-by-event ideal hydro simulations
  compared to the STAR data \cite{Adams:2004bi} for $Au+Au$ collisions at the RHIC energy $\sqrt{s}_{NN}=200$ GeV.}
  \label{fig:v2_pion_cent}
\end{figure}

\begin{figure}[t]
  \begin{center}
	\includegraphics[scale=0.45]{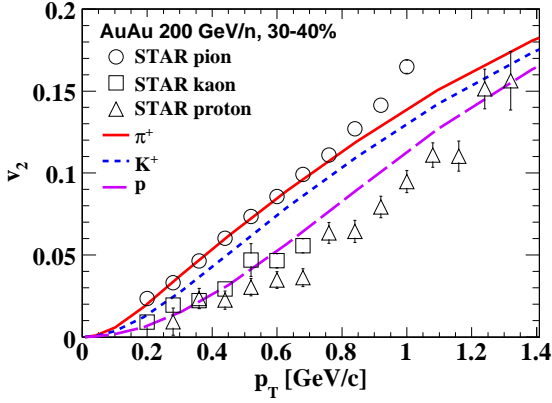}
  \end{center}
  \caption{(Color online) Elliptic flow for identified particles at centrality $30-40\%$ from event-by-event ideal hydro
   simulations (lines) compared to the  STAR data \cite{Adams:2004bi} (symbols).}
  \label{fig:v2_identified}
\end{figure}


To study the elliptic flow of identified hadrons, we show in Fig.~\ref{fig:v2_pion_cent} our ideal hydro results
on $v_{2}$ for positively charged pions in $Au+Au$ collisions at the RHIC energy $\sqrt{s}_{NN}=200$ GeV
that fit the STAR experimental data \cite{Adams:2004bi} very
well in $10-20\%$, $20-30\%$ and $30-40\%$ centrality bins for $p_{T}\leq 1$ GeV/$c$.
Such an agreement with experimental data in all centralities cannot be achieved with smoothed initial conditions
initial conditions according to Glauber model. One, however, can achieve the same agreement with smoothed initial condition 
but accounting for fluctuations in geometrical eccentricity  \cite{Song:2011hk}.
In our hydro calculations, we used $100$ events for each centrality bin. 
Much more events are needed for a true minimum bias ($0-80\%$) calculation using the 
AMPT initial conditions. For one single centrality bin $30-40\%$, our hydro results on $v_{2}$ 
for identified charged pions and anti-protons fit the STAR data  \cite{Adams:2004bi} reasonably well
as shown in Fig.~\ref{fig:v2_identified}. But the hydro results for charged kaons deviates significantly 
from the STAR data. 



\begin{figure}[t]
  \begin{center}
	\includegraphics[scale=0.45]{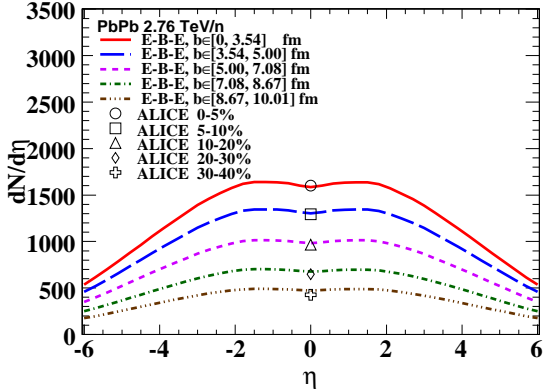}
  \end{center}
  \caption{(Color online) Charged hadron rapidity distributions from event-by-event ideal 
  hydro calculations in $Pb+Pb$ collisions at $\sqrt{s}=2.76$ TeV
  at different centralities as compared to the ALICE experimental data \cite{Aamodt:2010cz}.}
  \label{fig:dndeta-lhc}
\end{figure}

\begin{figure}[t]
  \begin{center}
	\includegraphics[scale=0.45]{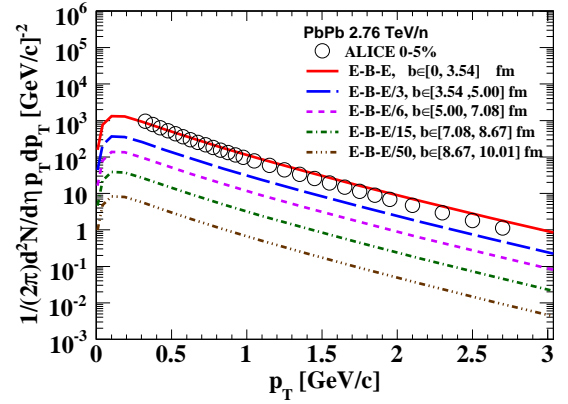}
  \end{center}
  \caption{(Color online) Transverse momentum spectra for charged hadrons in the central rapidity region $|\eta|<0.8$ from event-by-event
  ideal hydro  calculations in $Pb+Pb$ collisions at $\sqrt{s}=2.76$ TeV as compared to the ALICE data \cite{Aamodt:2010jd}.}
  \label{fig:dndpt-lhc}
\end{figure}

\begin{figure}[t]
  \begin{center}
	\includegraphics[scale=0.45]{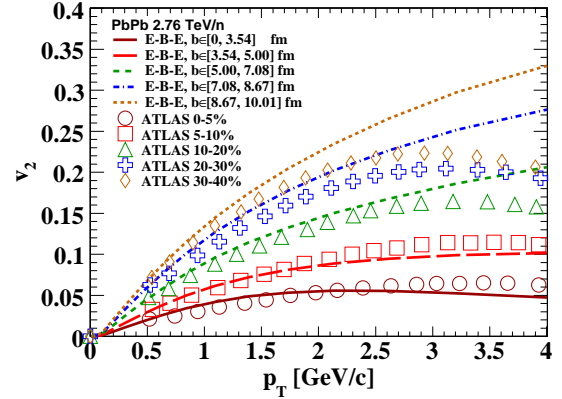}
  \end{center}
  \caption{(Color online) Elliptic flow of charged hadrons in the central rapidity region $|\eta|<2.5$ from event-by-event
  ideal hydro  calculations in $Pb+Pb$ collisions at $\sqrt{s}=2.76$ TeV compared to the ATLAS data \cite{Aad:2012bu}. 
  Event-planes are determined with charged hadrons in $3.3 <|\eta|<4.8$.}
  \label{fig:v2-lhc}
\end{figure}

\begin{figure}[t]
  \begin{center}
	\includegraphics[scale=0.5]{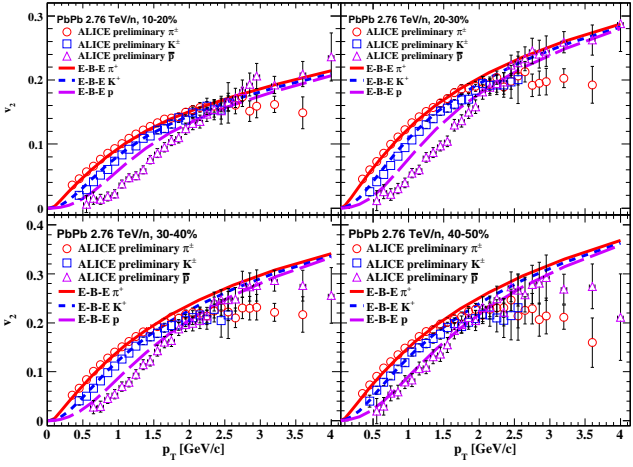}
  \end{center}
  \caption{(Color online) Elliptic flow from ideal hydro for identified particles in $Pb+Pb$ collisions 
  with four different centralities at the LHC
  energy $\sqrt{s}=2.76$  TeV compared to the preliminary ALICE data \cite{Collaboration:2011yba}.}
  \label{fig:v2lhc_identified}
\end{figure}

Following the same procedure with the same parameters for initial conditions from the AMPT model, we can predict hadron
spectra in $Pb+Pb$ collisions at the LHC energy $\sqrt{s}=2.76$ TeV. We have set the overall scale factor $K=1.6$ 
and the initial time $\tau_{0}=0.2$ fm/$c$ as constrained by the experimental data on the central rapidity density
of charged hadrons. Shown in Fig.~\ref{fig:dndeta-lhc} are the charged hadron
rapidity distributions in $Pb+Pb$ collisions at different centralities from our (3+1)D ideal hydro calculations.
The ranges of impact parameters are selected according to the centralities in the ALICE data \cite{Aamodt:2010cz} 
for the central rapidity region.

The corresponding transverse momentum spectra of charge hadrons in the central rapidity region $|\eta|<0.8$ 
for the most central $Pb+Pb$ collisions also
agree with the experimental data well as shown in Fig.~\ref{fig:dndpt-lhc}. The elliptic flow in $Pb+Pb$ collisions 
at the LHC energy $\sqrt{s}=2.76$ TeV in Fig.~\ref{fig:v2-lhc} is very similar to that in $Au+Au$ collisions 
at RHIC (see Fig.~\ref{fig:v2_EP_PP}). The ideal hydro results agree with the ALTAS experimental data well 
in central collisions but fail to describe the data at large $p_{T}$ in peripheral collisions, indicating the importance 
of viscous or non-equilibrium corrections. Here the charged
hadrons are restricted to $|\eta|<2.5$ and event-planes are determined with charged hadrons in the pseudo-rapidity window
$3.3 <|\eta|<4.8$ as in the ATLAS data. Shown in Fig.~\ref{fig:v2lhc_identified} are $v_{2}$ for identified hadrons as compared to the
preliminary ALICE data \cite{Collaboration:2011yba}. The (3+1)D hydro results describe the flavor dependence quite
well, except anti-protons in the 10-20\% centrality.

\section{Effects of longitudinal and flow velocity fluctuations}

\begin{figure}[t]
  \begin{center}
    \includegraphics[scale=0.6]{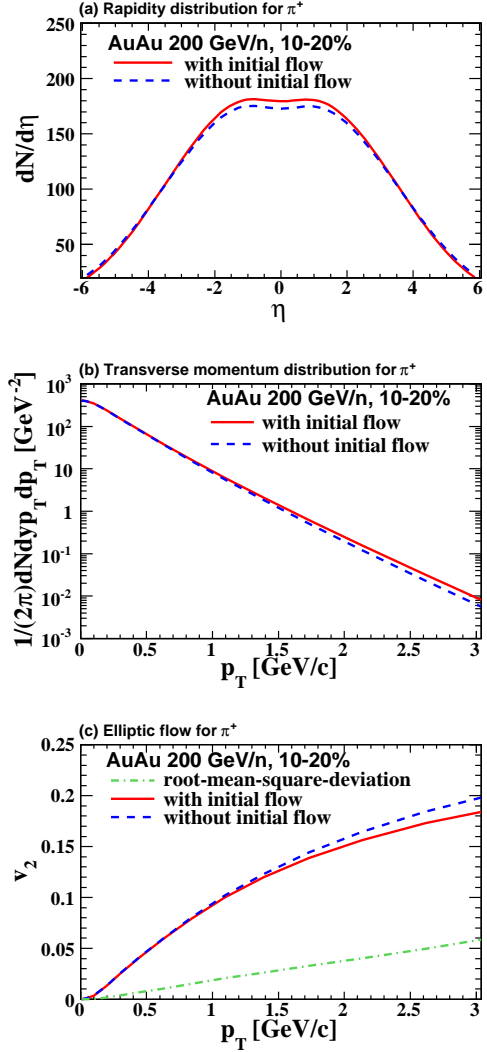}
  \end{center}
  \caption{(Color online) Event averaged multiplicity, $p_{T}$ spectra and $v_{2}$ comparison for WIF and WOF initial conditions at $10-20\%$ AuAu $200$ GeV/n collisions.}
  \label{fig:ini_flow}
\end{figure}

Most of the fluctuating initial conditions such as MC Glauber or MC KLN model assume zero initial transverse flow velocity
while the longitudinal flow velocity is assumed to be the same as the local spatial pseudo-rapidity $\eta_{s}$ in the Bjorken
scaling model. The latest (3+1)D viscous hydrodynamic model \cite{Schenke:2010rr} also assumes zero initial transverse
flow velocity and a Bjorken scaling scenario for the initial parton distribution in the longitudinal direction with an overall envelop function adjusted to reproduce the final hadron rapidity distribution. 

The initial condition in Eq.~(\ref{eq:Pmu}) from the AMPT model that we use in this study should contain 
non-vanishing local transverse and longitudinal flow velocities as well as fluctuations in the parton rapidity distribution.
The AMPT model uses the HIJING model for initial parton production which contains many mini-jets as well as excited strings. 
After initial thermalization via parton cascade within the initial time $\tau_{0}$ and the Gaussian smearing 
in  Eq.~(\ref{eq:Pmu}), these mini-jets will lead to small but non-vanishing collective radial flow velocities as well as 
large local flow velocity fluctuations. The local fluctuating initial flow velocities due to mini-jets should also have 
strong back-to-back correlation in azimuthal angle. Such initial collective radial flow and local velocity fluctuations 
should influence the final hadron spectra after hydrodynamic evolution.

To check the influence of the initial flow velocity fluctuations on hadron spectra, we show in
Fig.~\ref{fig:ini_flow} the pseudo-rapidity distributions (top), transverse momentum spectra (middle)
and elliptic flow of positively charged pions in 10-20\% central $Au+Au$ collision from our hydro
calculations with (solid) and without initial local flow velocities (dashed). These two initial conditions have the same initial
energy density distributions from AMPT simulations and the local flow velocities are set to zero in the latter case.
 There is a slight increase in the hadron multiplicity or initial total entropy and the slope of hadron spectra
and a slight decrease of the elliptic flow at high $p_{T}$ due to the fluctuating initial local flow velocities.

\begin{figure}[t]
  \begin{center}
    \includegraphics[scale=0.5]{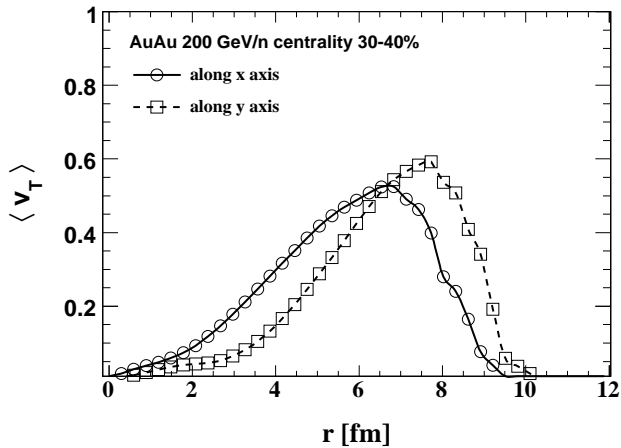}
  \end{center}
  \caption{ The event-averaged initial radial flow velocity along the $x$ and $y$-axis in the 30-40\% semi-central $Au+Au$ collisions
  at $\sqrt{s}=200$ GeV from the AMPT model at $\tau_{0}=0.4$ fm/$c$.}
  \label{fig:rflow}
\end{figure}

To understand the change in hadron transverse momentum spectra, we show in Fig.~\ref{fig:rflow} the
event-averaged initial radial flow velocities along the $x$ and $y$-axis in 30-40\% non-central $Au+Au$ 
collisions at $\sqrt{s}_{NN}=200$ GeV at an initial time $\tau_{0}=0.4$ fm/$c$.
One can see that parton interaction or thermalization during the initial time generates 
significant amount of radial flow that is anisotropic in azimuthal angle and is responsible for the harder transverse momentum 
spectra of final hadrons. Such radial flow velocities in the initial conditions for the hydrodynamic evolution
is also shown to influence the HBT correlation of final hadrons \cite{Pratt:2005bt}.

The fluctuation in the initial flow velocities seems to defuse the initial geometrical anisotropy a little
and leads to slightly smaller elliptic flow for final hadrons as shown in the lower panel of Fig.~\ref{fig:ini_flow}.
The calculated elliptic flow is determined mainly by the geometric eccentricity and averaged over many events. 
We find, however, there are significant differences between the elliptic flows
with and without the initial flow velocities on an event-by-event basis. We characterize these differences
by the variance
\begin{equation}
  \Delta v_{2} = \sqrt{\sum_{i=1}^{N_{\rm event}}(v_{2i}^{WIF}-v_{2i}^{WOF})^{2}/N_{\rm event}}
  \label{eq:rmsd}
\end{equation}
between the elliptic flow $v_{2}^{{WIF}}$ with the initial flow velocity and $v_{2}^{{WOF}}$ without, 
where $N_{\rm event}$  is the number of events. As shown by the dot-dashed line
in the lower panel of Fig.~ \ref{fig:ini_flow}, the variance is quite large reaching about 0.07 at $p_{T}=3$ GeV/$c$.
Note that we used hadrons in the rapidity range $3.1<|\eta|<3.9$ to determine the event-planes and calculate
the elliptic flow for hadrons in the central rapidity region  $|\eta|<1.0$.

\begin{figure}[t]
  \begin{center}
   \includegraphics[scale=0.5]{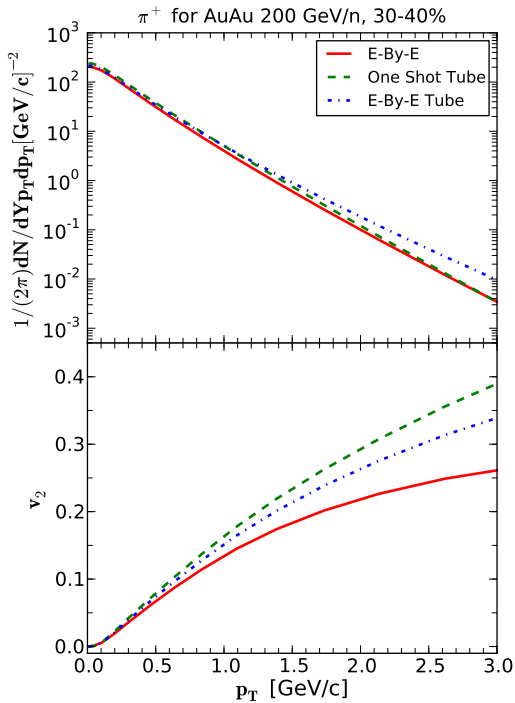}
  \end{center}
  \caption{ Transverse momentum spectra (upper panel) and elliptic flow (lower panel) for charged pions from hydrodynamic 
  simulations of 30-40\% semi-central $Au+Au$ collisions
  at $\sqrt{s}=200$ GeV with full fluctuating AMPT  initial conditions (solid lines), tube-like AMPT 
  initial conditions (dot-dashed lines) and one-shot AMPT tube-like initial conditions (dashed).}
  \label{fig:tube}
\end{figure}

To investigate the effect of longitudinal fluctuation on hadron spectra we compare our event-by-event
hydro calculations using the initial condition from the full AMPT results with that using a tube-like
smooth initial longitudinal distribution. In the tube-like initial condition, we take the initial energy-density
and transverse flow velocity from AMPT results in the central rapidity region and assume these
transverse fluctuations to be the same along the longitudinal direction with an envelop function
\begin{equation}
H(\eta)= \exp\left[-\theta(|\eta|-\eta_{0})(|\eta|-\eta_{0})^2/2\sigma_{w}^{2}\right],
\end{equation}
in rapidity. The length of the plateau $\eta_{0}$ in the central rapidity region and the width $\sigma_{w}$
of the Gaussian fall-off at large rapidities are adjusted to fit the charged hadron rapidity distribution.
As another comparison, we also calculate the elliptic flow from a one-shot AMPT initial condition, which
is the average of many AMPT events each rotated by an angle to a common participant-plane. We prefer this
as one-shot-tube initial condition since the initial parton density also has smooth tube-like distribution in the
longitudinal direction.
Shown in Fig.~\ref{fig:tube} are the transverse momentum spectra (upper panel) and elliptic flow (lower panel)
of charged pions in semi-central (30-40\%) $Au+Au$ collisions
at the RHIC energy $\sqrt{s}_{NN}=200$ GeV with the full AMPT initial conditions (solid lines) as compared to the 
initial conditions with a tube-like structure in the longitudinal direction (dot-dashed lines) and the one-shot 
AMPT with tube-like longitudinal distribution initial condition (dashed). The event-by-event 
fluctuations in the tube-like AMPT initial conditions
significantly reduce elliptic flow of final hadrons with respect to the event-planes as compared to the
one-shot-tube AMPT initial condition. The slope of the $p_{T}$ spectra from the event-by-event tube-like initial
conditions on the other hand is increased by the fluctuations (both the energy density and flow velocity) or 
hot spots in the transverse direction as compared to the spectra from one-shot-tube initial conditions.
Similar results were found by both (2+1)D \cite{Qiu:2011iv}
and (3+1)D hydro \cite{Schenke:2010rr} calculations. However, fluctuations in the longitudinal direction 
in the full AMPT initial conditions have also hot spots in the longitudinal direction. The expansion of 
such longitudinal hot spots will dissipate more transverse energy
into the longitudinal direction. This in turn decrease noticeably the value of the elliptic flow 
at large $p_{T}$ compared to the results from tube-like event-by-event AMPT initial conditions.
The slope of the $p_{T}$ spectra is also significant smaller than that from event-by-event tube-like AMPT
initial condition without fluctuation in the longitudinal direction.

Since anisotropic flow, at large $p_{T}$ in particular, is used to extract transport coefficients (such as shear viscosity) 
from comparisons between experimental data and viscous hydrodynamics, the inclusion of fluctuation 
in initial rapidity distribution in the hydrodynamic calculations will be necessary for more qualitative studies.

\begin{figure}[t]
  \begin{center}
    \includegraphics[scale=0.5]{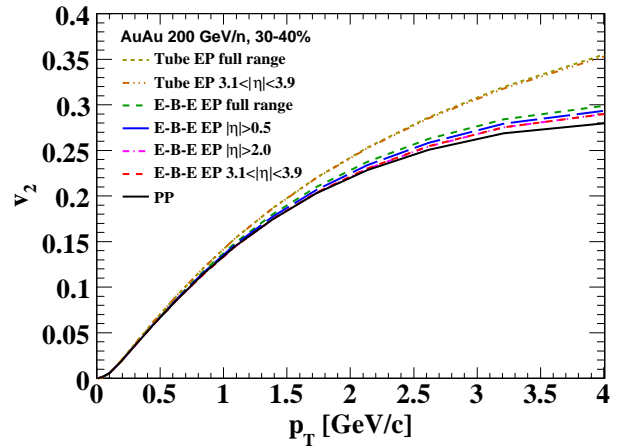}
  \end{center}
  \caption{Elliptic flows from (3+1)D hydro simulations of 30-40\% semi-central $Au+Au$ collisions
  at $\sqrt{s}=200$ GeV relative to the participant-planes (solid) and  event-planes determined by charged
  hadrons in the full rapidity window (dashed), $|\eta|>0.5$ (long-dashed), $|\eta|>2.0$ (dot-dashed),
  and $3.1<|\eta|<3.9$ (dotted).}
  \label{fig:eta-dep}
\end{figure}

In all the above calculations of elliptic flows for identified charged hadrons in the central rapidity region $|\eta|<1.0$,
the event-planes are determined using charged hadrons in $3.1<|\eta|<3.9$. To illustrate the sensitivity of the
calculated elliptic flows to the rapidity selection for charged hadrons that determine the event-plane, we show
in Fig.~\ref{fig:eta-dep} elliptic flows for hadrons in the central rapidity region with the event-planes determined
by hadrons in different rapidity widows. The dependence on the rapidity window is quite small. They are all slightly
larger than the elliptic flow measured against the participant-planes (solid lines).

Since mini-jets in initial conditions from the AMPT model contain both near-side and away-side
correlations, the fluctuations in the initial flow velocities that we use will also have important effects on final two-hadron 
correlations \cite{Pang:2009zm} in both azimuthal angle and rapidity. This is beyond the scope of our study in this paper and will be 
discussed in future studies.

\section{Partial Chemical Equilibrium}

Studies of hadron chemistry in high-energy heavy-ion collisions \cite{BraunMunzinger:2003zz} indicate that chemical 
equilibrium is reached during the early stage of the hadronic phase of the dense matter. The flavor abundance of the 
final hadrons indicates a freeze-out temperature $T_{cf}\approx 158-164$ MeV at the RHIC and LHC energies \cite{Andronic:2011yq}. 
This chemical freeze-out temperature is significantly higher than the kinetic freeze-out temperature $T_{f}=137$ MeV that
we used in our ideal hydro simulation. This is one of the reasons why the calculated proton spectra from our ideal
hydro simulations are about a factor of 1.2 lower than the experimental measurements as shown in Fig.~\ref{fig:dNdPt_identified}.

\begin{figure}[t]
  \begin{center}
    \includegraphics[scale=0.5]{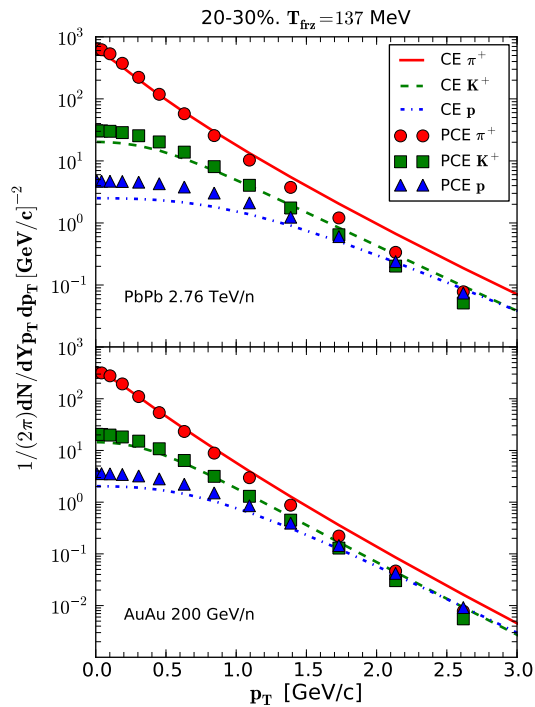}
  \end{center}
  \caption{Transverse momentum spectra of identified hadron spectra in semi-central (20-30\%) heavy-ion collisions at 
  RHIC (upper panel) and LHC (lower panel)
  energies from hydro simulations with CE (lines) and PCE (symbols) EoS.}
  \label{fig:pce-dndpt}
\end{figure}

\begin{figure}[t]
  \begin{center}
    \includegraphics[scale=0.5]{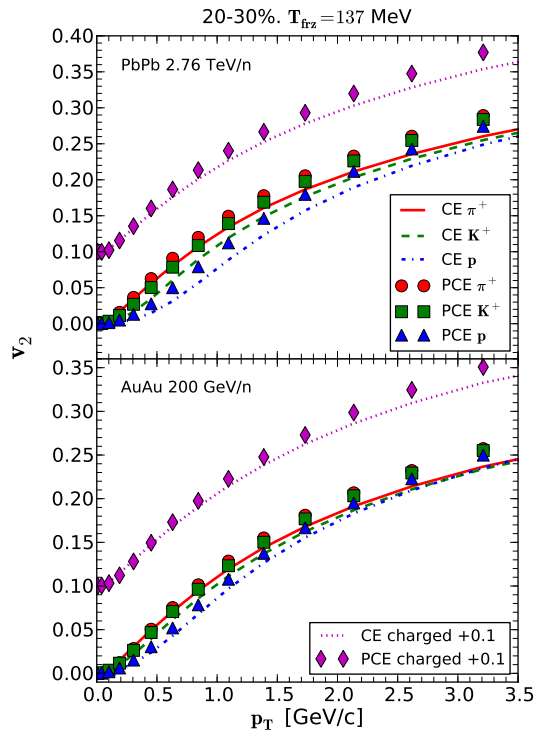}
  \end{center}
  \caption{Differential elliptic flow $v_{2}$ of identified and charged hadrons in semi-central (20-30\%) heavy-ion collisions at 
  RHIC (upper panel) and LHC (lower panel)
  energies from hydro simulations with CE (lines) and PCE (symbols) EoS. }
  \label{fig:pce-v2}
\end{figure}

To take into account the earlier chemical freeze-out during the hadronic phase of the hydrodynamical evolution, 
one should use a Partial Chemical Equilibrated (PCE) EoS \cite{techqm}.
We compare the hadron spectra and differential elliptic flow from hydro calculations with 
Chemical Equilibrium (CE) version s95p-v1 (lines) and PCE version s95p-PCE165-v0 EoS (symbols) at 
both RHIC and LHC energies in Figs.~\ref{fig:pce-dndpt} and \ref{fig:pce-v2}. We have used a kinetic freeze-out 
temperature $T_{frz}=137$ MeV in both calculations. Because of the finite chemical potential at the kinetic freeze-out
in the hydro with a PCE EoS, the corresponding kaon and proton yields at low $p_{T}$ is higher, improving agreement 
with the experimental data at RHIC. However, the slope of hadron spectra at high $p_{T}$ is steeper than that from hydro
with a CE EoS which is also below the experimental data. This can be improved a little by the viscous correction in the viscous
hydrodynamics. As shown in Fig.~\ref{fig:pce-v2}, the elliptic flow from hydro with a PCE EoS hydro is also about 20\%  
higher than that from hydro with a CE EoS, again leaving more rooms for improvement in the viscous hydrodynamics.
One should note that the PCE EoS'  are parameterized to take into account the higher chemical freeze-out temperature,
which however is different in heavy-ion collisions at different colliding energy \cite{Andronic:2011yq}. One therefore has to use
different parameterization of PCE EoS at different colliding energies. The PCE EoS that was fitted to RHIC data, however, cannot
describe the recent experimental data on identified hadron spectra from LHC using the viscous hydrodynamics \cite{Floris:2011ru}.

\section{Conclusion}

We have studied hadron spectra and elliptic flow in high-energy heavy-ion collisions within a (3+1)D ideal hydrodynamic model.
The (3+1)D ideal hydrodynamic equations are solved numerically using an extended FTC-SHASTA
algorithm. A projection method is developed to compute the freeze-out hyper surface for final hadrons.
We carried out event-by-event hydrodynamic simulations with fluctuating initial conditions for the energy-momentum
tensor from the AMPT model using a Gaussian smearing function for each initially produced parton. Such initial 
conditions provide both the local energy density as well as non-vanishing local flow velocities for the hydrodynamic evolution.
With a set of parameters (widths of the Gaussian smearing, an overall scale factor and initial time), the hadron rapidity distributions,
transverse momentum spectra as well as elliptic flows agree very well with experimental data, in particular at low $p_{T}$,
at both the RHIC and LHC energies.

We also illustrated the effects of local flow velocity in initial conditions from the AMPT model on final hadron spectra.
Due to rescattering during the initial time in the AMPT model, partons also develop some
initial collective radial flow which leads to harder transverse momentum spectra as compared to hydro calculations without initial
local flow velocities. The averaged elliptic flow however is not affected much. The fluctuation in the longitudinal distribution, however,
is shown to reduce the elliptic flow at large $p_{T}$ even in the central rapidity region. We also studied the sensitivities of the hydro 
results on the EoS and found influence of the early chemical freeze-out as parameterized in the PCE EoS to be important and should
be considered in hydro calculations at different colliding energies separately.

Our (3+1)D ideal hydro calculations provide an excellent description of hadron spectra at low $p_{T}$ in high-energy
heavy-ion collisions. At larger $p_{T}$, viscous corrections are known to become important \cite{Song:2007ux, Romatschke:2007mq}, in particular for higher hadronic flows. An extension of the study to the viscous hydrodynamics will be necessary for hadron
spectra at moderately large transverse momenta.

\section{APPENDIX: SHASTA algorithm}

There are high order and low order numerical algorithms to solve a partial differential equation. A low order algorithm keeps 
the solution monotonic but suffers from numerical diffusion, while a high order algorithm is more accurate but leads to 
dispersion (small ripples with new maximum and minimum during the evolution). The FCT (flux corrected transport) algorithm 
is developed to solve these problems in which the low order algorithm is used for the transport and diffusion, and a second step high order algorithm is used to do anti-diffusion with a corrected flux (equals to the diffusion term with ripples corrected). 
The SHASTA (SHarp And Smooth Transport Algorithm) algorithm is one kind of FCT algorithms which is good at dealing with 
strong gradients and shocks \cite{Boris:1973cp,Rischke:1995ir,Zalesak:1979}. We provide the basic steps of the FCT-SHASTA
algorithm in this appendix.

Hydrodynamic equations contain mainly partial differential equations whose basic form in one dimension is:
\begin{equation}
  \partial_{t}\rho + \partial_{x}(v\rho) = S(\rho,v),
  \label{eq:shasta_example}
\end{equation}
where $\rho$ can be mass density, energy density or momentum density and $v$ is the velocity along $x$ direction. 
In the following, we consider $\rho$ as the mass density for simplicity. The source term $S$ can be taken into 
account by a two-stage second-order mid-point Runge Kutta method: 

\begin{equation}
  \rho_{j}^{n+1} = \rho_{j}^{n}+\Delta t S_{j}^{1/2}(\rho_{j}^{n+1/2},v_{j}^{n+1/2}),
  \label{eq:Runge_Kutta}
\end{equation}
where $\rho_{j}^{n}$ denotes the mass density at grid point $j$ and time step $n$, $S_{j}^{1/2}$ are calculated from $\rho_{j}^{n+1/2}$ and $v_{j}^{n+1/2}$ given by the first stage half time step calculation in the Runge-Kutta method. To explain the SHASTA algorithm, we only consider $\partial_{t}\rho+\partial_{x}(v\rho)=0$ in this appendix. 

\subsection{The geometric interpretation of transport stage in SHASTA}

\begin{figure}[h]
  \begin{center}
	\includegraphics[scale=0.4]{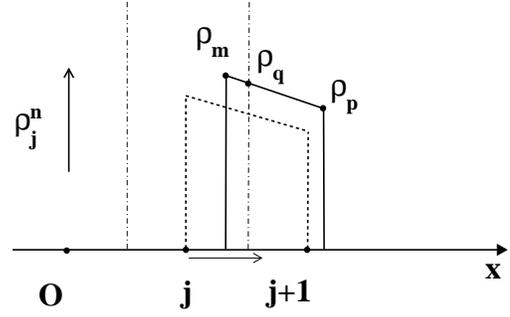}
  \end{center}
  \caption{The geometric explanation of SHASTA algorithm.  The dashed trapezoid represents the total mass in one fluid element between grid point $j$ and $j+1$ at time $t=0$. After one time step, the boundaries of this fluid element move to points $m$ and $p$ and the mass in this fluid element is conserved. One important principle of SHASTA algorithm is total mass conservation: $M=\Sigma_{j}\rho_{j}^{n}\delta x=\Sigma_{j}\rho_{j}^{n+1}\delta x$. The grid size $\delta x$ and time step $\delta t$ must satisfy $v\delta t<\delta x/2$ in the SHASTA algorithm to keep the positivity of the mass in each cell.}
  \label{fig:shasta}
\end{figure}

For the 1D FCT-SHASTA algorithm, we assume the velocity of the matter being transported at grid point $j$ between time $t$ and $t+\delta t$ 
can be approximated by $v_{j}^{1/2}$ at $t=t+\delta t/2$ which can be calculated from the 2-step Runge-Kutta method. The mass 
density $\rho_{j}$ and $\rho_{j+1}$ at the boundaries of one fluid element as illustrated in Fig.~\ref{fig:shasta} (the dashed trapezoid) 
change to $\rho_{m}$ and $\rho_{p}$ after one time step. Since the mass in this fluid element is conserved, 
\begin{equation}
  \frac{1}{2}(\rho_{j}^{0}+\rho_{j+1}^{0})\delta x=\frac{1}{2}(\rho_{m}+\rho_{p})
  \left[\delta x+(v_{j+1}^{1/2}-v_{j}^{1/2})\delta t\right].\\
  \label{eq:rho_p}
  \end{equation}
If we consider the two sides of the solid trapezoid vary in the same rate, we get 
  \begin{eqnarray}
	\rho_{p} & = & \rho_{j+1}^{0}\delta x/\left[\delta x+(v_{j+1}^{1/2}-v_{j}^{1/2})\delta t\right], \\
	\rho_{m} & = & \rho_{j}^{0}\delta x/\left[\delta x+(v_{j+1}^{1/2}-v_{j}^{1/2})\delta t\right].
	\end{eqnarray}
	
	Consider a cell centered at grid point $j$ (between the dot-dashed lines in Fig.~\ref{fig:shasta}). 
	The total mass in cell $j$ at time $t+\delta t$ 
comes from the mass moved in from cell $j-1$ to $j$ and the residual mass after some moved out from $j$ to $j+1$,
The mass density at point $q$ where the right boundary of cell $j$ intersects with the solid trapezoid is calculated from interpolation:	
	\begin{equation}
	\rho_{q}=\rho_{p}+(\rho_{m}-\rho_{p})(\delta x/2+v_{j+1}^{1/2}\delta t)/\left[\delta x+\delta t(v_{j+1}^{1/2}-v_{j}^{1/2})\right]
	  \label{eq:rho_q}
	\end{equation}
The residual mass is given by the area of the residual trapezoid:
	\begin{equation}
	  \begin{aligned}
	  \Delta m_{\rm re} & =  (\rho_{m}+\rho_{q})l_{\rm re}/2 \label{eq:mass_res} \\
	   & =  \delta x \left[ \frac{1}{2}Q_{+}^{2}(\rho_{j+1}^{n}-\rho_{j}^{n})+Q_{+}\rho_{j}^{n} \right],
	  \end{aligned}
	 \end{equation}
 where $l_{\rm re}=\delta x/2-v_{j}^{1/2}\delta t$ and 
	 \begin{eqnarray}
	  Q_{\pm}=(1/2 \mp v_{j}^{1/2}\delta t/\delta x)/\left[1 \pm (v_{j \pm1}^{1/2}- v_{j}^{1/2})\delta t/\delta x\right].
	   \label{eqn:QpQm}
	 \end{eqnarray}
	 Similarly, the mass transported from cell $j-1$ to cell $j$ can be calculated as,
	 	\begin{equation}
	  \Delta m_{\rm tr}  =  \delta x \left[ \frac{1}{2}Q_{-}^{2}(\rho_{j-1}^{n}-\rho_{j}^{n})+Q_{-}\rho_{j}^{n} \right].
	 \end{equation}
The mass density at grid point $j$ (averaged over cell $j$) and time step $n+1$ is therefore:
\begin{eqnarray}
	 \rho_{j}^{n+1} & =&  (\Delta m_{\rm re}+\Delta m_{\rm tr})/\delta x \nonumber \\
	  & =&  \frac{1}{2}Q_{-}^{2}(\rho_{j-1}^{0}-\rho_{j}^{0})+\frac{1}{2}Q_{+}^{2}(\rho_{j+1}^{0}-\rho_{j}^{0}) \nonumber \\ 
	  & & +(Q_{+}+Q_{-})\rho_{j}^{0} .
	   \label{eq:rho_new} 
\end{eqnarray}
For a uniform velocity, the above equation takes a simple form:
		  \[
		  \rho_{j}^{n+1}=\rho_{j}^{n}-\frac{\epsilon}{2}(\rho_{j+1}^{n}-\rho_{j-1}^{n})+(\frac{1}{8}+\frac{\epsilon^{2}}{2})(\rho_{j+1}^{n}-2\rho_{j}^{n}+\rho_{j-1}^{n})\]
		  where $\epsilon=v\delta t/\delta x$. For zero velocity, it becomes
		  \begin{equation}
		    \rho_{j}^{n+1}=\rho_{j}^{n}+\frac{1}{8}(\rho_{j+1}^{n}-2\rho_{j}^{n}+\rho_{j-1}^{n})
			\label{eq:trans_diffuse}
		  \end{equation}

 In the above transport stage of SHASTA, the solution is monotonic and positive but has a large zero order diffusion.

 \subsection{Anti-diffusion Stage}

 To correct the diffusion, an explicit form of anti-diffusion can be used (for the zero velocity case):
\begin{equation}
 \overline{\rho}_{j}^{n+1}=\rho_{j}^{n+1}-\frac{1}{8}(\rho_{j+1}^{n+1}-2\rho_{j}^{n+1}+\rho_{j-1}^{n+1}).
 \end{equation}

One can illustrate the above diffusion and anti-diffusion stage with a square wave initial field profile $(\cdots1,1,0,0,\cdots)$.
 The field becomes $(\cdots1,\frac{7}{8},\frac{1}{8},0,\cdots)$ after the transport stage, and 
 becomes $(\cdots\frac{65}{64},\frac{61}{64},\frac{3}{64},-\frac{1}{64},\cdots)$ after the explicit anti-diffusion stage. 
 One can see that new maximum and minimum are created, and the positivity is destroyed. 
 To solve this problem, the anti-diffusion terms are written in mass flux form:
\begin{equation}
 \bar{\rho}_{j}^{n+1}=\rho_{j}^{n+1}-f_{j+1/2}^{n+1}+f_{j-1/2}^{n+1},
 \end{equation}
 where the anti-diffusion mass flux is defined as:
\begin{equation}
 f_{j\pm1/2}^{n+1}=\pm\frac{1}{8}(\rho_{j\pm1}^{n+1}-\rho_{j}^{n+1}).
 \end{equation}
The above mass flux is further corrected as
		  \begin{equation}
		  f_{j+1/2}^{(c)n+1} = \sigma \max\left\{ 0, \min\left\{\sigma \Delta_{j-1/2},\frac{1}{8}\Delta_{j+1/2},\sigma \Delta_{j+3/2}\right\} \right\} 
			\label{eq:flux_corr}
		  \end{equation}
which is limited term by term so that no anti-diffusive-flux transfer of mass
can push the density at any grid point beyond the density value at neighboring points.
Here $\Delta_{j+1/2}=\rho_{j+1}^{n+1}-\rho_{j}^{n+1}$ and $\sigma=\text{sgn}\Delta_{j+1/2}$.
This is the origin of the name ``Flux Corrected Transport''.
The final mass density at grid $j$ and time step $n+1$ after corrected anti-diffusion stage is then
		  \begin{equation}
		   \bar{\rho}_{j}^{n+1}=\rho_{j}^{n+1}-f_{j+1/2}^{(c)n+1}+f_{j-1/2}^{(c)n+1},
			\label{eq:antidiff_corr}
		  \end{equation}
 where the diffusion is corrected by the anti-diffusion and the dispersion is corrected by the limitation on the mass flux. 
 The formula can be checked with several examples.

 In our (3+1)D hydrodynamics, we use a new limiter described by Zalesak \cite{Zalesak:1979} which is proved better than the 
 original one given by Boris and Book \cite{Boris:1973cp}. 

\section*{Acknowledgement}

We would like to thank helpful discussions with T. Hirano, B. Schenke and H. Song at various stages of this work, U. Heinz for
careful reading and useful comments on the manuscript. 
This work was supported by the Director, Office of Energy
Research, Office of High Energy and Nuclear Physics, Division of Nuclear Physics, of the U.S. Department of 
Energy under Contract No. DE-AC02-05CH11231 and within the framework of the JET Collaboration, 
by the National Natural Science Foundation of China
under the grant No. 11125524, and by self-determined research funds of CCNU from the collegesÕ basic 
research and operation of MOE.

\end{document}